\pdfoutput=1
\documentclass[11pt,reqno,preprint]{article}
\usepackage{jheppub,epsfig,amssymb,amsmath,mathrsfs,hyperref,array,hhline}
\usepackage{multirow,bbm}

\def\be{\begin{equation}}
\def\ee{\end{equation}}
\newcommand{\bea}{\begin{eqnarray}}
\newcommand{\eea}{\end{eqnarray}}

\def\Fig#1{Fig.~{\ref{#1}}}

\def\sect#1{section~{\ref{#1}}}
\def\eqn#1{eq.~(\ref{#1})}

\def\Tab#1{Table~{\ref{#1}}}

\def\ket#1{\langle #1 \rangle}
\def\lr{\leftrightarrow}

\def\Gcusp{\Gamma_{\rm cusp}}
\def\z{{\zeta}}
\def\Li{\textrm{Li}}
\def\eps{\epsilon}

\newcommand{\cA}{\begin{cal}A\end{cal}}
\newcommand{\cB}{\begin{cal}B\end{cal}}

\newcommand{\cE}{\begin{cal}E\end{cal}}

\newcommand{\cN}{\begin{cal}N\end{cal}}

\newcommand{\cP}{\begin{cal}P\end{cal}}

\newcommand{\cS}{\begin{cal}S\end{cal}}

\newcommand{\cZ}{\begin{cal}Z\end{cal}}

\def\g#1{{\color{green}#1}}

\def\red#1{{\color{red}#1}}

\title{Lifting Heptagon Symbols to Functions}

\author{Lance~J.~Dixon$^{1}$ and Yu-Ting~Liu$^{1}$}

\affiliation{$^1$ SLAC National Accelerator Laboratory,
Stanford University, Stanford, CA 94309, USA}

\abstract{Seven-point amplitudes in planar ${\cal N}=4$ super-Yang-Mills
theory have previously been constructed through four loops using the
Steinmann cluster bootstrap, but only at the level of the symbol.
We promote these symbols to actual functions, by specifying their
first derivatives and boundary conditions on a particular two-dimensional
surface. To do this, we impose branch-cut conditions and
construct the entire heptagon function space through weight six.
We plot the amplitudes on a few lines in the bulk Euclidean region,
and explore the properties of the heptagon function space under
the coaction associated with multiple polylogarithms.}

\emailAdd{lance@slac.stanford.edu}
\emailAdd{aytliu@stanford.edu}

\preprint{ \begin{flushright} SLAC--PUB--17544 \end{flushright}}

\begin{document}
\maketitle
\flushbottom

%%%%%%%%%%%%%%%%%%%%%%%%%%%%%%%%%%%%%%%%%%%%%%%%%%%%%%%%%%%

\section{Introduction}
The study of scattering amplitudes from their analytic properties has a long history which dates back to the beginning of the $S$-matrix program (see e.g.~ref.~\cite{ELOP}).  One recent incarnation of this program imposes fairly general constraints on scattering (typically $2\to2$ scattering), and leads, for example in two spacetime dimensions, to bounds on couplings and other parameters which are often satisfied by particular known theories~\cite{Paulos:2016but,EliasMiro:2019kyf,Cordova:2019lot,Bercini:2019vme}. In contrast, the planar $\cN=4$ super-Yang-Mills (SYM) amplitude bootstrap program starts with a fixed theory, the simplest gauge theory in four dimensions, and aims to compute arbitrary scattering amplitudes, typically in perturbation theory but without ever directly inspecting the loop integrand.  Instead, one uses a knowledge (or suspicion) about the space of functions to which the amplitudes belong, including their branch-cut behavior, some properties of their first derivatives, and their behavior in soft, collinear, and/or multi-Regge kinematical limits.  (For a recent review, see ref.~\cite{Caron-Huot:2020bkp}.)

Planar $\cN=4$ SYM exhibits a number of remarkable properties that make it a suitable playground for developing and exploiting novel computational techniques. In particular, it possesses a dual superconformal symmetry~\cite{Drummond:2006rz,Bern:2006ew,Bern:2007ct,Alday:2007hr,Alday:2007he,Drummond:2007au,Drummond:2008vq,CaronHuot:2011ky} in addition to the usual superconformal symmetry. The dual superconformal symmetry eliminates all kinematical degrees of freedom at four and five points, where the amplitudes are uniquely fixed by their infrared divergences, as captured by the Bern-Dixon-Smirnov (BDS) ansatz~\cite{Bern:2005iz}. Starting at six points, the BDS ansatz for $n$-point amplitudes receives infrared-finite corrections~\cite{Drummond:2007bm,Bartels:2008ce,Bern:2008ap,Drummond:2008aq}, which depend only on $3(n-5)$ independent dual conformal cross ratios.  The correction to the maximally helicity violating (MHV) amplitude has traditionally been expressed in terms of a remainder function~\cite{Bern:2008ap,Drummond:2008aq,Dixon:2011pw,Dixon:2013eka,Drummond:2014ffa}, and the correction to the next-to-maximally helicity violating (NMHV) amplitude in terms of the NMHV ratio function~\cite{Drummond:2008vq,Drummond:2008bq,Kosower:2010yk,Dixon:2011nj,Dixon:2014iba,Dixon:2015iva}.

Dual conformal symmetry allows an alternative description of the kinematic space in terms of \emph{momentum twistors}~\cite{Hodges:2009hk}, which make this symmetry manifest.  It is expected~\cite{ArkaniHamed:2012nw} that MHV and NMHV amplitudes in planar $\cN=4$ SYM at arbitrary loop order can be expressed in terms of a class of transcendental functions called multiple polylogarithms~\cite{Gonch3} (modulo a pinch of salt~\cite{Brown:2020rda}).  These functions are graded by an integer \emph{weight}, the number of integrations, and endowed with a Hopf algebra coaction~\cite{Gonch3,FBThesis,Duhr:2011zq,Duhr:2012fh}, whose maximal iteration is called the \emph{symbol}~\cite{Goncharov:2010jf}. A link has been observed~\cite{Golden:2013xva} between the arguments of transcendental functions, encoded as the \emph{entries} or \emph{letters} of the symbol, that appear  and the $\cA$-coordinates of certain types of cluster algebras~\cite{Fomin2001-bo,Fomin2002-pw}. At six and seven points, the corresponding cluster algebras are of finite type; this feature allows one to bootstrap the amplitudes by writing a general ansatz as a linear combination of all possible functions having the appropriate symbol alphabet of 9 and 42 letters, respectively, multiplied by unknown rational-number coefficients. One can solve for the coefficients in the ansatz by using different mathematical and physical constraints.  To be more specific, in perturbation theory the $L$-loop remainder and ratio functions are expected to be multiple polylogarithms of weight $2L$ whose symbol letters are drawn from cluster $\cA$-coordinates.

The amplitude bootstrap becomes much more efficient when one only considers functions with branch cuts in the correct locations, which constrains the first entry of the symbol~\cite{Gaiotto:2011dt}. It can be streamlined even further by imposing the Steinmann relations~\cite{Steinmann,Steinmann2} forbidding overlapping branch cuts~\cite{Caron-Huot:2016owq}, which constrain the first two symbol entries.  There are further constraints deeper into the symbol, which can be described physically as extended Steinmann relations~\cite{Caron-Huot:2018dsv,Caron-Huot:2019bsq} or mathematically as cluster adjacency conditions~\cite{Drummond:2017ssj,Drummond:2018dfd,Drummond:2018caf,Golden:2019kks,Gurdogan:2020tip,Mago:2020eua}. Recently, patterns in the symbols of seven- and higher-point amplitudes have been associated with tropicalizations of the associated Grassmannian~\cite{Drummond:2019qjk,Drummond:2019cxm,Arkani-Hamed:2019rds,Henke:2019hve,Drummond:2020kqg}.

To date, the six-point amplitudes in planar $\cN=4$ SYM have been computed to seven loops~\cite{Caron-Huot:2019vjl,DDtoappear}. These results are available at the level of full functions, making it possible to plot the results, in principle in any kinematics, as well as to explore the analytic structure in many different limits. The seven-point amplitudes have been bootstrapped to four loops~\cite{CaronHuot:2011kk,Drummond:2014ffa,Dixon:2016nkn,Drummond:2018caf}, but only at the level of the symbol. Recall that the symbol captures the iterated branch cut structure of a function, but it omits information about constants at every step of the integration. (See ref.~\cite{Duhr:2012fh} for a review and application to physics.)  On the other hand, the symbol information can be a very important computational springboard for computing the full function~\cite{Dixon:2013eka}.  As full multiple polylogarithmic functions, the seven point amplitude is only known to two loops and only for the MHV amplitude~\cite{Golden:2014xqf,Golden:2018gtk,Bourjaily:2019vby}, for which the symbol (actually the total differential) was found earlier~\cite{CaronHuot:2011ky}.

There are several motivations for lifting the symbols of amplitudes to functions. First, numerical values of the amplitudes can be used to study properties of the theory itself. For example, using the six-point amplitudes obtained from the bootstrap in planar $\cN=4$ SYM, the ratio of amplitudes at successive loop orders, evaluated for generic kinematics, tends toward a constant~\cite{Dixon:2014voa,Dixon:2015iva,Caron-Huot:2019vjl}, which is a signature of the finite radius of convergence of the theory. In the same papers, perturbative results for the six-point remainder function were also found to have strikingly similar behavior to the strong coupling results obtained from the AdS/CFT correspondence~\cite{Alday:2007hr,Alday:2009dv}.  Such numerical information is simply not available from the symbol.

Furthermore, in many interesting kinematic regions, important analytic information is buried in the \emph{beyond-the-symbol} terms of the amplitudes. For example, amplitudes in multi-Regge kinematics (MRK) have been predicted to all orders in the six-point case~\cite{Basso:2014pla} and they match the limiting behavior of the bootstrapped amplitudes through seven loops~\cite{Caron-Huot:2019vjl,DDtoappear}, an incredibly powerful test.  More recently, an all-orders proposal for the multi-Regge limit of arbitrary $n$-point amplitudes has been presented~\cite{DelDuca:2019tur}. (See however ref.~\cite{Bartels:2020twc}, which argues that new Reggeon cuts open up first at eight points.)  The new ingredient appearing in ref.~\cite{DelDuca:2019tur} is the central emission block, which first appears in the seven-point amplitude.  It would be interesting to check the predictions for the central emission block using bootstrapped amplitudes at the level of full functions.

Another example of an interesting kinematic region is the ``origin'', which for the six-point case entails taking all three cross-ratios to zero. In this case, perturbative data~\cite{Caron-Huot:2019vjl} and all-orders arguments~\cite{Basso:2020xts} show that the logarithm of the MHV amplitude depends only quadratically on logarithms of the cross ratios.  Beyond one loop, the quadratic dependence gets multiplied by transcendental constants (in this case, Riemann zeta values), which can all be expressed in terms of a ``tilted'' version~\cite{Basso:2020xts} of the Beisert-Eden-Staudacher (BES) kernel controlling the cusp anomalous dimension at finite coupling~\cite{Beisert:2006ez}.   Such zeta values are completely invisible at the symbol level.  Therefore, studying analogous kinematic regions at higher multiplicities requires full knowledge of the amplitudes as functions.

Finally, the information contained in beyond-the-symbol functions, and specifically the transcendental constants, highly enriches the study of the \emph{cosmic Galois coaction principle}.  In the space of hexagon functions, this principle organizes and implies certain restrictions in the space of zeta values required as independent functions, and the zeta values appearing at a particular base point when all cross ratios are equal to unity~\cite{Caron-Huot:2019bsq}.  Similar restrictions on higher-loop behavior from lower-loop results have been seen earlier for primitive divergences in $\phi^4$ theory~\cite{Schnetz:2013hqa,Panzer:2016snt,Brown:2015fyf} and for the electron anomalous magnetic moment~\cite{Schnetz:2017bko}.  It would be interesting to study the coaction principle in planar $\cN=4$ SYM for seven particles, and its relation to the coaction principle for six particles, since the kinematics are smoothly connected through soft and collinear limits.

In this paper, we bridge the gap between symbols and functions at seven points, by constructing a complete space of heptagon functions through weight six.  Then we lift the known symbols of the amplitudes to functions within this space through four loops.  (At four loops, we characterize the functions by their ``double derivatives'', more precisely, their $\{6,1,1\}$ coproducts, which lie in the weight six space.)  The key step, defining the heptagon functions, requires imposing various branch-cut conditions, which are the analog of the first-entry and Steinmann conditions discussed earlier. For the six-point case, various aspects of this procedure have been discussed in several places, e.g.~refs.~\cite{Dixon:2013eka,Caron-Huot:2016owq,Caron-Huot:2019bsq,Caron-Huot:2020bkp}.  However, our seven-point implementation will be somewhat different.

The six-point case has a distinguished kinematical point in the ``bulk'', where all three cross ratios are unity.  This point is invariant under all dihedral transformations of the hexagon, and all hexagon functions are finite there and evaluate to multiple zeta values (MZVs).  Furthermore, this point is connected to the soft and MRK limits of the six-point kinematics by lines on which the functions evaluate to harmonic polylogarithms (HPLs) depending on a single variable~\cite{Remiddi:1999ew}, making it simple to ``get around'' in the space.  We know of no such distinguished bulk point in the seven-point case. (There is a dihedrally invariant point in the bulk, but functions seem unlikely to be very simple there.) Instead, we will impose the branch cut conditions and fix the amplitudes solely on the boundary of the kinematics.  Our workhorse will be a two-dimensional surface that can be defined as a triple-scaling limit, in which four of the six independent cross ratios are infinitesimal, and the other two are generic.  We call this the ``CO'' surface because it interpolates between collinear (C) kinematics and a seven-point origin (O).  Functions on this surface are still very simple; they are just logarithms in the small cross ratios, and one-variable HPLs in each of the two generic ones.  Yet the surface is still rich enough to allow us to impose an almost complete set of branch-cut conditions, as well as to touch many interesting kinematic regions, and thereby easily move around information such as constants of integration.

Just as in the six-point case, the branch cut conditions couple together, through derivatives, or more precisely the coaction, functions which have nonvanishing symbols, and beyond-the-symbol functions which have the form of MZVs multiplied by lower-weight functions.  We will find that no such functions are required using the first such zeta value, $\zeta_2 = \pi^2/6$. This situation is similar to what happens in the six-point case.  However, at weight 3 we find that $\zeta_3$ is an independent element of the function space (unlike the six-point case), and it spawns a tower of independent functions multiplied by $\zeta_3$.  Once the function space is constructed, we determine the amplitudes within it using their limiting behavior, as well as final-entry conditions arising from dual superconformal invariance~\cite{Bullimore:2011kg,CaronHuot:2011kk,Dixon:2016nkn}.  As was found earlier at symbol level~\cite{Drummond:2014ffa,Dixon:2016nkn,Drummond:2018caf}, fewer types of constraints are required to determine seven-point amplitudes than were needed for six-point amplitudes; we will need to use only soft limits and vanishing of spurious poles to fix the coefficients of the beyond-the-symbol functions.
% \lance{Double check this assertion.}

Once we have fixed all the amplitudes, essentially by characterizing their first derivatives iteratively, and providing boundary conditions on the CO surface, we can integrate them up off this surface.  We do so for three separate lines emanating from the CO surface, and plot the results for successive loop order ratios.  We also discuss briefly the amplitudes' behavior at the seven-point origin, reserving a more detailed examination of the MHV amplitude in that region to another publication~\cite{BDLPtoappear}.

This paper is organized as follows. In \sect{sec:review} we review the basic properties of seven-point amplitudes in planar $\cN=4$ SYM and the Steinmann cluster bootstrap. In \sect{sec:liftingsymbols} we define the Collinear-Origin surface and explain how to use it, in conjunction with branch cut conditions, to define heptagon functions.  Then in \sect{sec:liftingsymbols} we determine the MHV and NMHV amplitudes through four loops in terms of these functions. Next, in \sect{sec:bulk} we plot the amplitudes on three lines in the bulk, and in \sect{sec:origin} we provide some information about their behavior at the origin.  In \sect{sec:coactioncomments} we comment on what coaction-like restrictions can be seen so far in the heptagon functions, and on the function-level validity of various symbol-level constraints arising from tropical fans. Finally, in \sect{sec:conclusions} we conclude and provide an outlook for further research directions enabled by our results.

We provide several computer-readabled ancillary files along with this paper:
The file \texttt{R\_P\_o\_co.txt} gives the MHV remainder function
and NMHV ratio function at the origin and on the CO surface.
\texttt{HCoproductTables.txt} defines the complete space of
heptagon functions via their $\{n-1,1\}$ coproducts through weight 6.
The weight 7 and weight 8 functions needed to describe the four-loop
amplitude components are defined similarly in the files
\texttt{MHE\_7.txt}, \texttt{MHO\_7.txt}, 
\texttt{MHE\_8.txt}, and \texttt{MHO\_8.txt}.
The representation of the amplitudes in terms of heptagon functions
is provided in \texttt{AmpsH.txt}.  The action of the generators
of the dihedral group $D_7$ on all the heptagon functions is given in
\texttt{HDihedralSym.txt}, while \texttt{HcoTable.txt} presents their
values on the CO surface.  Finally, \texttt{weight6odd406.txt}
identifies a 406-dimensional subspace of the 412 weight 6 parity-odd
functions that is singled out by the amplitude coproducts.
The files are too large to accompany an arXiv submission or journal
article, so they are hosted at~\cite{HepfnsWebsite}.

%%%%%%%%%%%%%%%%%%%%%%%%%%%%%%%%%%%%%%%%%%%%%%%%%%%%%%%%%%%%%%%%%

\section{Review and notation}
\label{sec:review}

\subsection{Kinematics, dual conformal cross ratios and momentum twistors}

In planar $\cN=4$ SYM, amplitudes are known to respect a dual (super)conformal symmetry, which is a conformal symmetry in the dual coordinates $x_i$,
whose differences are the momenta of the $n$ external particles,
\be 
p_i = x_{i+1} - x_i,
\ee
along with the dual of the fermionic supermomenta, whose definition we do not need here.  Momentum conservation becomes trivial in the dual coordinates once we identify the points $x_{i+n} \equiv x_i$. Due to the dual conformal symmetry, the infrared-finite part of the scattering amplitudes only depends on the dual conformally invariant cross ratios,
\be
u_{i,j} = \frac{x_{i,j+1}^2\ x_{i+1,j}^2}{x_{i,j}^2\ x_{i+1,j+1}^2} \ ,
\ee
where $x_{i,j} \equiv x_i - x_j$.

Because of the on-shell conditions $x_{i,i+1}^2 = 0$, for $n=4,5$ all the cross ratios are trivial. As a consequence, the amplitudes are completely fixed by the BDS ansatz~\cite{Bern:2005iz}. When $n \ge 6$, the amplitudes become dependent on the non-trivial cross ratios. For seven particles, we need the seven cross ratios
\be\label{eq:uidef}
u_i \equiv u_{i+1,i+4} \,,
\ee
which are given in terms of the 2- and 3-particle Mandelstam invariants $s_{i,i+1} = (p_i+p_{i+1})^2$ and $s_{i,i+1,i+2} = (p_i+p_{i+1}+p_{i+2})^2$ in \eqn{eq:uisi}.  Only six of the seven $u_i$ are independent, due to a Gram determinant condition,
\bea\label{eq:gram}
0 = 1 + \biggl[-u_1 + u_1 u_3 &+& u_1 u_4 + u_1 u_2 u_5 - u_1 u_3 u_5 - u_1^2 u_4 u_5 - 2 u_1 u_2 u_4 u_5 \nonumber\\
  &+& u_1 u_2 u_3 u_5 u_6 + u_1^2 u_2 u_4 u_5^2 \ +\ \text{cyclic}\biggr] + u_1 u_2 u_3 u_4 u_5 u_6 u_7 \,,
\eea
where a cyclic transformation takes $u_i$ to $u_{i+1}$ modulo 7, and `$+$~cyclic' in the equation above means summing over all 7 images of the expression in square brackets under cyclic rotations.

Dual (super)conformal symmetry also makes it possible to describe the kinematics with \emph{momentum (super)twistors}~\cite{Hodges:2009hk,Mason:2009qx},
\be
\cZ_i = (Z_i \, | \, \chi_i),
\ee
where $Z_i\in \mathbb{P}^3$ are the bosonic momentum twistors and $\chi_i$ their fermionic counterparts.  Here we only summarize what we need in this paper; see e.g.~ref.~\cite{Elvang:2013cua} for their definitions and a pedagogical review.

Invariant quantities are all constructed from momentum twistor four-brackets,
$\left<ijkl\right> = \det(Z_i Z_j Z_k Z_l)$.  A subset of the four-brackets are related to the dual coordinates,
\be
x_{i,j}^2
= \frac{\left<i-1,i,j-1,j\right>}{\left<i-1,i\right> \left<j-1,j\right>} \,,
\ee
where $\left<ij\right> = \det(\lambda_i\lambda_j)$ is the usual spinor product.

We also need to define the dual superconformal $R$-invariant, or five-bracket,
\be\label{fivebrak}
[abcde] = \frac{\delta^{0|4} \big( \chi_a \langle b c d e \rangle + \text{cyclic} \big)}{\langle a b c d \rangle \langle b c d e \rangle \langle c d e a \rangle \langle d e a b \rangle \langle e a b c \rangle} \,,
\ee
where again `$+$~cyclic' means summing over all 5 cyclic rotations generated by
$a\to b\to c\to d\to e\to a$.  The $R$-invariants show up in NMHV amplitudes and they obey the six-term identity,
\be\label{eq:sixZidentity}
[abcde] - [bcdef] + [cdefa] - [defab] + [efabc] - [fabcd] = 0\,.
\ee
For seven particles, we adopt the notation of ref.~\cite{Dixon:2016nkn} and write the five-bracket in terms of the two omitted labels,
\be
(67) = (76) \equiv [12345], \quad\quad \text{etc.}
\ee

%%%%%%%%%%%%%%%%%%%%%%%%%%%%%%

\subsection{Seven-particle amplitudes and BDS(-like) normalizations}

For a gauge theory, the color-ordered partial amplitudes $A_n$ for $n$ particle scattering are the coefficients of the single traces ${\text Tr}(T^{a_1}T^{a_2}\cdots T^{a_n})$ in the color decomposition. In the planar limit, they completely characterize the full-color scattering amplitudes. For $\cN=4$ SYM, the particle and helicity content can be packaged into a superfield,
\be
\Phi=G^+ +\eta^A\Gamma_A+\tfrac{1}{2!}\eta^A\eta^B S_{AB}+\tfrac{1}{3!}\eta^A\eta^B\eta^C\epsilon_{ABCD}\bar \Gamma^D+\tfrac{1}{4!}\eta^A\eta^B\eta^C\eta^D\epsilon_{ABCD}G^-\,,
\ee
for each leg, with auxiliary Grassmann variables $\eta^A$, where $A\in\{1,2,3,4\}$ is the SU(4) $R$-symmetry index. The amplitudes for different particles and/or helicities are then packaged into a \emph{superamplitude} $\cA_n(\Phi_i)$, $i=1,2,\ldots n$.  The expansion of $\cA_n$ in the Grassmann variables naturally organizes it into components in an N$^k$MHV expansion,
\be\label{eq:NkMHVexpansion}
\cA_n = \cA_n^{\text{MHV}} + \cA_n^{\text{NMHV}} + \cA_n^{\text{NNMHV}} + \dots + \cA_n^{\overline{\text{MHV}}} \,.
\ee

For seven particles, we only need to know $\cA_7^{\text{MHV}}$ and $\cA_7^{\text{NMHV}}$; the remaining two components in \eqn{eq:NkMHVexpansion} are related by parity conjugation. Furthermore, we can express the correction to the BDS ansatz (see appendix~\ref{app:bds}) in terms of the \emph{remainder function} $R_n$,
\be
\cA_n^{\text{MHV}} = \cA_n^{\text{BDS}} \exp(R_n) \,,
\ee
and the \emph{ratio function} $\cP_n$,
\be
\cA_n^{\text{NMHV}} = \cA_n^{\text{MHV}} \cP_n.
\ee
The information contained in the remainder and ratio functions is equivalent to that in the \emph{BDS-normalized} amplitudes, defined by
\begin{align}
\cB_n &\equiv \frac{\cA_n^{\text{MHV}}}{\cA_n^{\text{BDS}}} = \exp(R_n) \,,\\
B_n &\equiv \frac{\cA_n^{\text{NMHV}}}{\cA_n^{\text{BDS}}} = \cP_n \, \cB_n \, .
\end{align}

Working in perturbation theory, any quantity $F$ related to the amplitudes can be expanded as
\be\label{eq:gLoopExpansion}
F=\sum_{L=0}^\infty g^{2L} F^{(L)} \,,
\ee
where $g^2 = {g_{\text{YM}}^{2}N_c}/{(16 \pi^{2})}$, $g_{\text{YM}}$ is the Yang-Mills coupling constant, $N_c$ is the number of colors, and $F^{(L)}$ is the $L$-loop contribution to $F$.  Specifically, the cusp anomalous dimension $\Gcusp$, known to
all orders in planar $\cN=4$ SYM~\cite{Beisert:2006ez}, is expanded as
\be\label{eq:Gcusp}
\frac{\Gcusp}{4} = g^2 - 2\,\z_2\,g^4 + 22\,\z_4\,g^6 - \left(219\,\z_6 + 8\,\z_3^2\right) g^8 + \mathcal{O}(g^{10})\,.
\ee

As discussed in ref.~\cite{Dixon:2016nkn} (following the analysis for six particles~\cite{Caron-Huot:2016owq}), it is convenient to normalize the amplitudes differently, in order to remove non-trivial dependence on three-particle Mandelstam variables from $\cA_n^{\text{BDS}}$. We define a \emph{BDS-like} ansatz~\cite{Alday:2009dv},
\be\label{eq:bdslike}
\cA_n^{\text{BDS-like}} \equiv \cA_n^{\text{BDS}}
\exp\left[ -\frac{\Gcusp}{4} \, \cE_n^{(1)} \right] \,,
\ee
where in our case $n=7$,
\be
\cE_7^{(1)} = \sum_{i=1}^7 \biggl[ \Li_2\left(1-\frac{1}{u_i}\right)
  + \frac{1}{2} \ln \left(\frac{u_{i+2}u_{i{-}2}}{u_{i+3}u_{i}u_{i{-}3}}\right)
               \ln u_i \biggr].
\ee
The purpose of normalizing by the BDS-like ansatz is to preserve the Steinmann relations, which forbid overlapping three-particle cuts, in the dual-conformally invariant \emph{BDS-like normalized} amplitudes, defined by
\begin{align}
\cE_n &\equiv \frac{\cA_n^{\text{MHV}}}{\cA_n^{\text{BDS-like}}}
= \exp\left[R_n + \frac{\Gcusp}{4} \, \cE_n^{(1)}\right]
= \cB_n \, \exp\left[\frac{\Gcusp}{4} \, \cE_n^{(1)}\right] \,, \\
E_n &\equiv \frac{\cA_n^{\text{NMHV}}}{\cA_n^{\text{BDS-like}}} = \cP_n \, \cE_n
= B_n \, \exp\left[\frac{\Gcusp}{4} \, \cE_n^{(1)}\right] \,.
\end{align}
Because $R_n$ starts at two loops, by using the expansion of $\Gcusp$, \eqn{eq:Gcusp}, we see that indeed $\cE_n^{(1)}$ is the one-loop BDS-like-normalized amplitude. From now on, we will focus on seven-particle amplitudes and drop the subscript $n=7$.

The NMHV amplitude is a sum of bosonic functions multiplied by $R$-invariants. Due to the six-term identity~(\ref{eq:sixZidentity}), there are only 15 independent $R$-invariants. We choose them to be $(12)$, $(14)$, plus their cyclic images, and $\cP^{(0)}$, where
\be\label{eq:P7tree}
\cP^{(0)} = \frac{3}{7} \, (12) + \frac{1}{7} \, (13) + \frac{2}{7} \, (14) ~ + \text{cyclic}
\ee
is the tree-level ratio function~\cite{Drummond:2008bq}. We can now write the NMHV amplitude as
\begin{align}
B &= \cP^{(0)} \, B_0 + \big[ (12) \, B_{12} + (14) \, B_{14} + \text{ cyclic} \big]\,, & &\text{(BDS-normalized)}
\label{eq:B7components}\\
E &= \cP^{(0)} \, E_0 + \big[ (12) \, E_{12} + (14) \, E_{14} + \text{ cyclic} \big]\,. & &\text{(BDS-like-normalized)}
\label{eq:E7components}
\end{align}
We will refer to the bosonic functions $B_0$, $E_0$, $B_{12}$, $E_{12}$, etc., as components of the NMHV amplitude.

%%%%%%%%%%%%%%%%%%%%%%%%%%%%%%

\subsection{The Steinmann cluster bootstrap}

We recall that a multiple polylogarithm (MPL) $F$ of weight $n$ is an iterated integral, which can be defined recursively through its total differential,
\be\label{eq:mpldiff}
dF = \sum_\phi F^{\phi}\ d\ln \phi,
\ee
where each $\phi$ is a symbol letter. The collection of all symbol letters is the \emph{symbol alphabet}. $F^{\phi}$ is a weight $n-1$ MPL which we will call the ($\{n-1,1\}$-)\emph{coproduct} of $F$ with respect to $\phi$. Since logarithms of products are additive, we see that we only need to consider a multiplicatively independent set of symbol letters. It is often convenient to have explicit representations of MPLs (see e.g.~those in ref.~\cite{Duhr:2012fh}). Again these are defined recursively, for a weight $n$ MPL, by
\be\label{eq:gfunc}
G(z_1,z_2,\dots,z_n; z) \equiv \int_0^z \frac{dt}{t-z_1} G(z_2,\dots,z_n; t),
\ee
with the weight $0$ case defined to be $G(; z) \equiv 1$. The special case where $z_1,\dots,z_n$ are all 0 is defined to be
\be
G(0,\dots,0; z) \equiv \frac{1}{n!}\ln^n z.
\ee
We call the above representation of MPLs the \emph{$G$-functions}. We will also use the obvious notation $G(\vec{w}; z)$ where $\vec{w} = (w_1,\dots,w_n)$ is the weight vector.

The symbol of an MPL is also defined recursively,
\be\label{eq:symboldiff}
\cS(F) = \sum_\phi \cS(F^{\phi}) \otimes \phi\,,
\ee
summing over the symbol alphabet. Specifically, the symbol of a weight $n$ MPL is a linear combination of tensor products of $n$ letters, and the coefficient of each term is a rational number.  There are also constants in the MPL space with nontrivial weight, typically multiple zeta values. By definition, the symbol does not `see' the constants,
\be
\cS(\text{MZV}) = 0.
\ee
Therefore if we wish to integrate up a symbol using \eqn{eq:mpldiff}, an ambiguity shows up at every step of the integration. That is, the symbol does not uniquely specify the function to which it is associated.

In the (Steinmann) cluster bootstrap, we assume that the symbol alphabet is drawn from the $\cA$-coordinates of certain cluster algebras~\cite{Golden:2013xva}. The $\cA$-coordinates can be expressed in terms of the Plücker coordinates, or four-brackets of the momentum twistors already introduced above. For the seven-point case, a multiplicatively independent, projectively invariant basis of 42 symbol letters is given by~\cite{Drummond:2014ffa},
\begin{align}
\label{aletters}
a_{11} &= \frac{\ket{1234}\ket{1567}\ket{2367}}{\ket{1237}\ket{1267}\ket{3456}}\,, \quad & \quad
a_{41} &= \frac{\ket{2457}\ket{3456}}{\ket{2345}\ket{4567}}\,, \nonumber \\
a_{21} &= \frac{\ket{1234}\ket{2567}}{\ket{1267}\ket{2345}}\,, \quad & \quad
a_{51} &= \frac{\ket{1(23)(45)(67)}}{\ket{1234}\ket{1567}}\,, \\
a_{31} &= \frac{\ket{1567}\ket{2347}}{\ket{1237}\ket{4567}}\,, \quad & \quad
a_{61} &= \frac{\ket{1(34)(56)(72)}}{\ket{1234}\ket{1567}}\,, \nonumber
\end{align}
plus their cyclic permutations $a_{ij} \equiv a_{i1}\big |_{Z_{k} \to Z_{k+j-1}}$, $j=1,2,\ldots,7$. In the above expressions, the Plücker bilinear is defined as
\be
\ket{a(bc)(de)(fg)} \equiv \ket{abde}\ket{acfg}-\ket{abfg}\ket{acde} \,.
\ee

In addition to the alphabet basis~(\ref{aletters}), it is convenient to introduce another equivalent basis, related to the above $a$-basis by\footnote{We thank G.~Papathanasiou and A.~McLeod for discussions of the $g$-basis.}
\begin{align}
\label{gletters}
g_{11} &= \frac{a_{17}}{a_{13}a_{14}}\,, \quad & \quad
g_{41} &= \frac{a_{47}a_{57}}{a_{12}a_{15}a_{17}}\,, \nonumber \\
g_{21} &= \frac{a_{24}a_{33}}{a_{13}a_{14}}\,, \quad & \quad
g_{51} &= \frac{a_{24}}{a_{33}}\,, \\
g_{31} &= \frac{a_{67}}{a_{11}a_{16}}\,, \quad & \quad
g_{61} &= \frac{a_{47}}{a_{57}}\,, \nonumber
\end{align}
along with their cyclic permutations $g_{ij} \equiv g_{i1}\big |_{Z_{k} \to Z_{k+j-1}}$, for $j=1,2,\ldots,7$. The $g$-basis is arranged to have ascending complexity when written in terms of the cross ratios $u_i$. For example, the first four cyclic orbits of the $g$-letters are,
\begin{align*}  \label{eq:glettersu}
g_{11} &= u_1\,, \quad & \quad
g_{31} &= 1 - u_3u_6\,,  \\
g_{21} &= 1 - u_1\,, \quad & \quad
g_{41} &= 1 - u_2u_5 - u_4u_7\,. 
\end{align*}
Furthermore, the $g$-letters have definite parity: $g_{5i}$ and $g_{6i}$ are parity odd (and thus involve square roots when written in terms of the $u_i$), while the rest are parity even,
\begin{align}
\text{Parity transformation:}
\begin{cases}
g_{ki} \to 1/g_{ki} \quad &\text{for } k = 5,6\\
g_{ki} \to g_{ki} \quad &\text{otherwise.}
\end{cases}
\end{align}
We also record here the action of the generators of the dihedral group $D_7$,
\begin{align}
\text{Cyclic transformation: }&
g_{k,i} \to g_{k,i+1} \\
\text{Flip transformation: }&
\begin{cases}
g_{k,i} \to 1/g_{k,8-i} &\text{if } k = 5 \\
g_{k,i} \to g_{k,8-i} &\text{otherwise.}
\end{cases}
\end{align}

Given the symbol alphabet, one can build iteratively a \emph{function space} at each weight, which satisfies certain constraints one expects the amplitudes to satisfy. These constraints include physical branch cuts, integrability, (extended) Steinmann relations~\cite{Steinmann,Steinmann2,Caron-Huot:2016owq,Dixon:2016nkn,Caron-Huot:2018dsv,Caron-Huot:2019bsq}, and cluster adjacency~\cite{Drummond:2017ssj,Drummond:2018dfd}. At the end of the day one writes the amplitudes at a given loop order $L$ as a linear combination of weight $2L$ functions in the function space, each function multiplied by an unknown rational-number coefficient. One can then fix these unknown coefficients with additional constraints such as the final-entry condition~\cite{Bullimore:2011kg,CaronHuot:2011kk} and the known behavior in soft or collinear limits. For details of this procedure at symbol level, we refer the reader to previous work~\cite{Drummond:2014ffa,Dixon:2016nkn,Drummond:2018caf} which bootstrapped the symbols of the amplitudes through four loops.  In the next section we will focus on the new ingredients necessary at function level.

%%%%%%%%%%%%%%%%%%%%%%%%%%%%%%%%%%%%%%%%%%%%%%%

\section{Lifting symbols to functions}
\label{sec:liftingsymbols}

\subsection{Boundary of integration --- the Collinear-Origin surface}
\label{sec:COsubsection}

In order to lift symbols to functions, we will again work our way up weight by weight through integration.  Or to say it differently, we will define weight $n$ functions iteratively through their $\{n-1,1\}$ coproducts~(\ref{eq:mpldiff}), but in contrast to the symbol definition~(\ref{eq:symboldiff}), we also have to specify the coefficients of functions that vanish at symbol level, because they have zeta values multiplying them.  These coefficients will largely be fixed by physical branch cut conditions.  We also need to provide constants of integration somewhere in the kinematical space.  As mentioned in the introduction, we will use a particular surface on the boundary of the kinematical space, where several cross ratios are infinitesimal.

\begin{figure}[h]
\center
\includegraphics[width=0.5\linewidth]{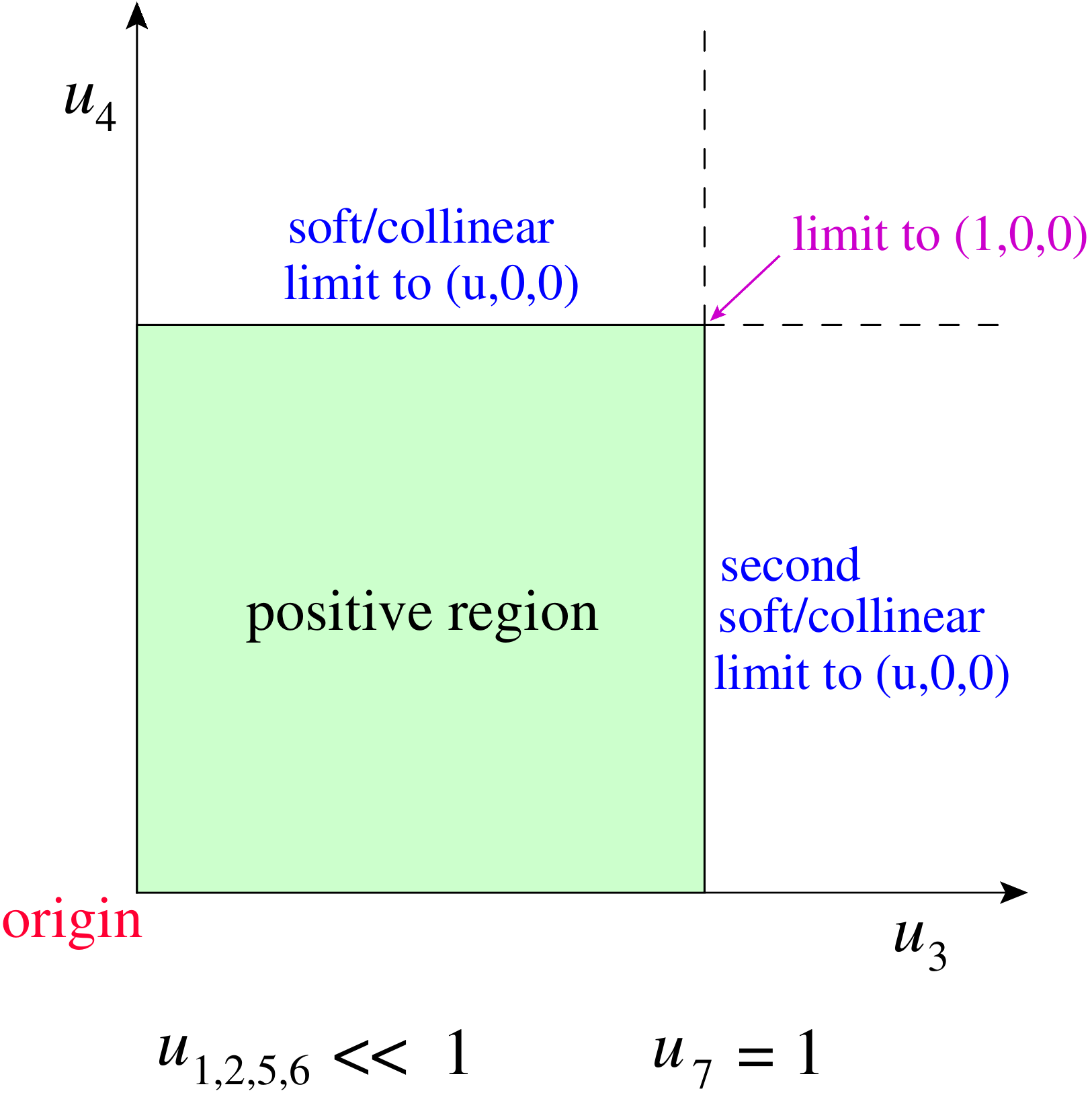}
\caption{The plane in $u_3$, $u_4$ defining the CO surface.}
\label{Fig:COsurface}
\end{figure}

We define the \emph{Collinear-Origin} (CO) surface, which interpolates between the heptagon \emph{origin} and the soft/collinear limits. The heptagon origin is the seven-point analog of the hexagon origin \cite{Basso:2020xts}, where as many of the cross ratios $u_i$ go to zero as possible. (The $u_i$ at six and seven points all contain two-particle invariants in their numerator, and so the origin can be defined by maximizing how many two-particle invariants vanish.) At seven points, six of the cross ratios (say $u_1$ to $u_6$) become infinitesimal, while the last one ($u_7$) goes to unity due to the Gram determinant constraint~(\ref{eq:gram}). The CO surface has the following kinematics,
\be
% \text{\bf{Collinear-origin:}} \quad
u_1,u_2,u_5,u_6 \ll 1, \qquad\ u_7 \to 1,
\ee
while $u_3$ and $u_4$ are generic. It is depicted in \Fig{Fig:COsurface}.  The origin is the limit $(u_3,u_4)\to (0,0)$. The limits $u_3\to 1$ and $u_4\to 1$ are soft limits onto six-point kinematics where two of the three six-point cross ratios (say $v,w$) are infinitesimal; we refer to this as the $(u,0,0)$ line.  (See \sect{sec:MHVamp} for more details.)  The intersection of these two lines is a double soft limit onto the six-point kinematics $(u,v,w)=(1,0,0)$, and on another Riemann sheet it corresponds to a multi-Regge limit. The intersection of the positive region, in which all cluster coordinates are positive, with the CO surface is the unit square $0 < u_3,u_4 < 1$ and is highlighted in green.

One can reach the CO surface using a triple scaling limit from a specific momentum twistor parametrization. (Indeed, this is how it was identified.)  We use the parametrization described in \cite{Basso:2013aha} in the context of the Pentagon Operator Product Expansion, and define the variables,
\be
T_j = e^{-\tau_j}, \qquad
S_j = e^{\sigma_j}, \qquad
F_j = e^{i\phi_j},
\ee
for $j = 1,2$, so that the momentum twistors become,
\be\label{eq:TSF7}
(Z_1,Z_2,\ldots,Z_7) =
\begin{pmatrix}
\frac{S_1}{\sqrt{F_1}}  & 1 & -1 & -S_2\sqrt{F_2} & 0 & 0 & 0 \\
0 & 0 & 0 & \frac{1}{T_2\sqrt{F_2}} & \frac{S_2+T_2F_2}{T_2S_2\sqrt{F_2}}
        & 1 & \frac{1}{S_1\sqrt{F_1}} \\
\frac{\sqrt{F_1}}{T_1} & 0 & 0 & -\frac{1}{T_2\sqrt{F_2}}
        & -\frac{1}{T_2\sqrt{F_2}} & 0 & \frac{\sqrt{F_1}}{T_1} \\
T_1\sqrt{F_1} & 0 & 1 & \frac{1+T_2S_2F_2+T_2^2}{T_2\sqrt{F_2}}
        & \frac{1}{T_2\sqrt{F_2}} & 0 & 0
\end{pmatrix} \,.
\ee
We then take the limit
\be\label{eq:tslimit}
T_j \mapsto T_j \cdot \eps,  \qquad
S_j \mapsto S_j \cdot \eps^{-1}, \qquad
F_j \mapsto F_j \cdot \eps^{-2} 
\ee
with $\eps \to 0$. In this limit, using \eqn{gletters} to write $g_{1i}=u_i$ in terms of the $a$-letters, and the four-bracket representation~(\ref{aletters}) of the $a$-letters, the cross ratios become
\begin{align}\label{eq:tsui}
&u_1 \to \frac{\eps^2}{S_2 (S_2 + F_2 T_2)}\,, & \quad &u_5\to \eps^2\cdot T_1^2\,, \nonumber\\
&u_2 \to \eps^2\cdot T_2^2\,, & \quad &u_6\to \frac{\eps^2}{S_1 (S_1 + F_1 T_1)}\,, \nonumber\\
&u_3 \to \frac{S_2}{S_2 + F_2 T_2}\,, & \quad &u_7\to 1\,. \\
&u_4 \to \frac{S_1}{S_1 + F_1 T_1}\,, \nonumber
\end{align}
where indeed $u_1,u_2,u_5,u_6$ are infinitesimal.

Furthermore, on the CO surface, the original 42 symbol letters collapse to a simple alphabet of 9 letters (see \eqn{eq:COgletters} in appendix~\ref{app:symco}),
\be\label{eq:coletters}
u_1, \quad u_2, \quad u_3, \quad 1-u_3, \quad u_4, \quad 1-u_4, \quad u_5, \quad u_6, \quad 1-u_7,
\ee
and they are parametrized by the six cross ratios $u_1$ through $u_6$. The last cross ratio $u_7$ is related to the others by the Gram determinant constraint~(\ref{eq:gram}), which reduces in this limit to,
\be\label{eq:COGramdet}
1-u_7 = u_1(1-u_4) + u_6(1-u_3).
\ee
As part of the branch cut conditions, which we will discuss further in \sect{sec:functionConstraints}, the coproduct with respect to $(1-u_7)$ vanishes everywhere on the CO surface. This vanishing can already be checked at the symbol level from the symbols provided in the ancillary files for ref.~\cite{Dixon:2016nkn}.

As a result of this vanishing, only the first 8 letters in \eqn{eq:coletters} are effective, and the functions factorize as products of functions that depend on each $u_i$ only, for $i = 1,2,\ldots,6$. To be specific, $u_1,u_2,u_5,u_6$ only have singularities at 0; therefore they integrate trivially to powers of logarithms $\ln^k(u_i)$. For $u_3$ and $u_4$, there are singularities at 0 and 1, and they integrate to $G$-functions $G(\vec{w};u_i)$ with $\vec{w}$ a vector consisting of 0's and 1's only; traditionally these functions are called \emph{harmonic polylogarithms} or HPLs~\cite{Remiddi:1999ew}, $H_{\vec{w}}(u_i)$, which are equivalent to $G(\vec{w};u_i)$ up to a minus-sign convention. We also allow for transcendental constants, which we assume to be drawn from multiple zeta values (MZVs). Overall, heptagon functions on the CO surface are contained within the following factorized function space,
\be\label{eq:COfunctions}
\{1,\z_2,\z_3,\dots \} \otimes
\left( \bigotimes_{i\in\{1,2,5,6\}}\left\{\ln^k(u_i),\ k\ge0\right\} \right) \otimes
\left( \bigotimes_{i\in\{3,4\}}\left\{H_{\vec{w}}(u_i),\ w_k\in\{0,1\} \right\} \right).
\ee

The CO surface allows us to implement the full dihedral symmetry of the bulk. First of all, it is symmetric under the flip that leaves $u_7$ invariant,
\be\label{eq:COflip}
u_1 \lr u_6, \quad u_2 \lr u_5, \quad u_3 \lr u_4\,.
\ee
This symmetry is equivalent to exchanging the two sets of $F_j$, $S_j$ and $T_j$ in the operator product expansion (OPE) parametrization.  One can see explicitly how this flip symmetry acts on the $g$-letters in appendix~\ref{app:symco}.

Secondly, and more importantly, the point $\{u_3 \to 0, u_4 \to 1\}$ on the surface is related to the point $\{u_3 \to 1, u_4 \to 0\}$ through cycling
\be\label{eq:COcycle}
u_i \to u_{i+3} \,,
\ee
plus a parity transformation.  The parity transformation is invisible in the alphabet (\ref{eq:coletters}); an extra minus sign is needed for each parity-odd function.  The cyclic symmetry~(\ref{eq:COcycle}) means that the seven different cyclic images of the CO surface touch at points.  That is enough ``contact'' to be able to transport constants of integration from one CO surface to another, and fully implement any dihedral symmetry constraints.

%%%%%%%%%%%%%%%%%%%%%%%

\subsection{Lifting the coproduct table}

In the symbol-level bootstrap, the function space at each weight is often represented~\cite{Dixon:2013eka,Dixon:2015iva,Dixon:2016nkn} through a \emph{coproduct table} $c_{ij\phi}$,
\be\label{eq:symcop}
\cS(F^{(n)}_i) = \sum_{\phi,j} c_{ij\phi}\ \cS(F^{(n-1)}_j) \otimes \phi\,,
\ee
where $F^{(n)}_i$ is the $i^{\rm th}$ function at weight $n$; the sum above is over all symbol letters $\phi$ and all functions $F^{(n-1)}_j$ in the weight $(n-1)$ function space.  Note that in this equation, we use the superscript (in parentheses) to denote the weight of the function, whereas in \eqn{eq:mpldiff} we use the superscript (without parentheses) to denote the coproduct with respect to the symbol letter $\phi$; the meaning of the superscript should always be clear from the context.  The tensor $c_{ij\phi}$ has dimension $s_n \times s_{n-1} \times |\Phi|$, where $s_n$ is the dimension of the space of symbols at weight $n$, and $|\Phi|$ is the number of symbol letters (42 in our case).  This nested or iterative representation of the function space is more economical than expanding out all the symbols; for example, the symbol of the BDS-like normalized four-loop MHV amplitude has 105,403,942 terms in the $a$-letters~\cite{Dixon:2016nkn}.  It also generalizes easily to a function-level description, as at six points~\cite{Dixon:2013eka}.

To obtain the functions as MPLs, the symbol level coproduct table (\ref{eq:symcop}) tells us how to integrate,
\be\label{eq:fncop}
dF^{(n)}_i = \sum \hat{c}_{ij\phi}\ F^{(n-1)}_j d\ln\phi,
\ee
but with an ambiguity; that is, we now have to include functions $F^{(n-1)}_j$ that have vanishing symbols, such as transcendental constants, or transcendental constants multiplied by lower weight MPLs.  The coefficients multiplying these \emph{beyond-the-symbol} functions on the right-hand side are not yet known and need to be fixed.

In other words, $\hat{c}_{ij\phi}$ has dimension $(s_n+z_n) \times (s_{n-1}+z_{n-1}) \times |\Phi|$, where $z_n$ is the dimension of the space of beyond-the-symbol functions at weight $n$.  For the $s_n \times s_{n-1} \times |\Phi|$ dimensional sub-tensor we can re-use the symbol-level table, $\hat{c}_{ij\phi} = c_{ij\phi}$ in this case.  Also, the $z_n \times s_{n-1} \times |\Phi|$ dimensional sub-tensor vanishes identically, because the weight $(n-1)$ function produced by coacting on a function with a zeta value multiplying it will also contain that zeta value as a factor. Related to this point, the $z_n \times z_{n-1} \times |\Phi|$ dimensional sub-tensor can be copied from a lower-weight case, by multiplying the lower-weight functions by appropriate zeta values.  The nontrivial part is the $s_n \times z_{n-1} \times |\Phi|$ dimensional sub-tensor.

%%%%%%%%%%%%%%%%%%%%%%%%%%%%%%%%
\subsection{Constraints for beyond-the-symbol terms}
\label{sec:functionConstraints}

To fix the beyond-the-symbol terms in the coproduct table, we impose several constraints. Many of them are already discussed in detail in previous work,
e.g.~refs.~\cite{Dixon:2013eka,Caron-Huot:2019bsq,Dixon:2016nkn}.

\emph{Dihedral symmetry and parity.} \quad
We would like the functions to respect the same symmetry as their symbols. For example, if the symbols of two functions $F_1$, $F_2$ are related by a cyclic transformation,
\be
\cS(F_2) = \cS(F_1) \Big|_{Z_{k}\to Z_{k+1}}\,,
\nonumber
\ee
then we require the functions themselves to be related by the same transformation,
\be
F_2 = F_1 \Big|_{Z_{k}\to Z_{k+1}}\,,
\nonumber
\ee
and similarly for the flip and parity transformations.

\emph{Integrability.} \quad
This is the condition that partial derivatives should commute, or in a concise notation,
\be
d^2F = 0\,,
\ee
with the derivative $d$ given by \eqn{eq:fncop}.  This condition tends to be the most computationally demanding one to impose.  However, in our case we can leverage the fact that integrability was already solved previously at symbol level~\cite{Dixon:2016nkn}, and so we do not need to solve nearly as large a system of equations for the much smaller set of beyond-the-symbol unknowns ($z_n \ll s_n$).

\emph{Branch cuts.} \quad
For a physical massless scattering amplitude, branch points can only occur when Mandelstam invariants vanish, $x_{i,j}^2 = 0$. In terms of the cross ratios, the branch points (on the Euclidean sheet) can only be located at $u_i = 0$ or $u_i = \infty$ for some $u_i$. At the symbol level, this gives rise to the \emph{first-entry condition}~\cite{Gaiotto:2011dt}, which says that the first entry (or the left-most entry, in our notation) of the symbol can only be one of the 7 cross ratios $u_i$. At the function level, the branch cut condition means that a function $F$ should not have a logarithmic dependence on $\phi$ when a non-first-entry letter $\phi$ approaches zero.  To be more precise, suppose we have an underlying kinematic variable $x$ such that $\phi$ depends on $x$ near $\phi=0$, $\partial\phi/\partial x|_{\phi=0} \neq 0$, but all other letters that depend on $x$ (call them $\psi(x)$) are nonvanishing as $\phi\to0$.  Then
\be
\frac{\partial F}{\partial x}\biggr|_{\phi\to0}
= \frac{\partial\phi}{\partial x}\biggr|_{\phi\to0} \frac{F^\phi}{\phi}
  \ +\ \text{nonsingular.}
\ee
To avoid the logarithmic singularity, we impose the coproduct condition
\be\label{eq:Fphieq0}
\left. F^{\phi} \right\vert_{\phi\to0,\ \psi\not\to0 \text{ for all other }\psi(x)} = 0\,,
\ee
or in terms of the coproduct table entries,
\be
\sum_j \hat{c}_{ij\phi} \left. F_j^{(n-1)}
\right\vert_{\phi\to0,\ \psi\not\to0 \text{ for all other }\psi(x)} = 0
\ee
for $\phi \notin \{u_k\}$.

Because of the complicated dependence of the letters on the underlying
kinematic variables, \eqn{eq:Fphieq0} could be tricky to impose at an
arbitrary point.  However, it simplifies on the CO surface, where
we can take $u_1$ to $u_6$ to be independent variables.
Equation~(\ref{eq:COgletters}) gives the limiting behavior of the $g$-letters
on the CO surface.  Together with \eqn{eq:mpldiff} and the chain rule,
we see that on the CO surface
\be\label{eq:COu3deriv}
\frac{\partial F}{\partial u_3} =
\frac{F^{u_3}}{u_3} - \frac{F^{1-u_3}}{1-u_3} \,,
\ee
where
\bea
F^{u_3} &=& F^{g_{13}} - F^{g_{51}} + F^{g_{55}} - F^{g_{56}} + F^{g_{57}}
- F^{g_{61}} + F^{g_{62}} - F^{g_{63}} + F^{g_{64}} - F^{g_{65}} \,,
\label{eq:Fu3} \\
F^{1-u_3} &=& F^{g_{23}} + F^{g_{35}} + F^{g_{44}} + F^{g_{46}}
- F^{g_{53}} + 2 \, F^{g_{54}} - 2 \, F^{g_{55}}
+ 2 \, F^{g_{56}} - 2 \, F^{g_{57}}  \nonumber\\
&&\hskip0.1cm\null
+ 2 \, F^{g_{61}} - 2 \, F^{g_{62}} + 2 \, F^{g_{63}} - F^{g_{64}}
+ F^{g_{66}} - 2 \, F^{g_{67}} \,.
\label{eq:Fomu3}
\eea
To avoid a logarithmic singularity at $u_3=1$, we need $F^{1-u_3}$
to vanish there.  The flip symmetry~(\ref{eq:COflip}) of the CO surface
implies a similar condition on $F^{1-u_4}$. Finally, we consider a derivative
with respect to $u_7$, evaluated on the CO surface ($u_7=1$) for generic
$u_3$ and $u_4$.  (In this case we use \eqn{eq:COGramdet} to trade say $u_4$
for $u_7$, so that $u_4$ depends on $u_7$, and we assume that
neither $u_4$ nor $1-u_4$ vanishes.)

In summary, we have the three conditions,
\be\label{eq:CObranchcut}
\begin{split}
F^{1-u_3} = 0 \quad & \quad \text{when } u_3 \to 1\,, \\
F^{1-u_4} = 0 \quad & \quad \text{when } u_4 \to 1\,, \\
F^{1-u_7} = 0 \quad & \quad \text{on the whole CO surface.} 
\end{split}
\ee
These quantities refer to coproducts computed on the CO surface,
in terms of the variables parametrizing the surface (or in the case
of $1-u_7$, a small departure from the surface).  We actually want
to impose constraints on the ``bulk'' coproducts $F^{g_{ij}}$, so that
they define the derivatives of functions globally in the kinematics.
For $F^{1-u_3}$, we use \eqn{eq:Fomu3}; for $F^{1-u_4}$, the same equation
after applying the flip symmetry~(\ref{eq:COflip}); and for $F^{1-u_7}$
we find from \eqn{eq:COgletters} that
\be\label{eq:Fomu7} 
F^{1-u_7} = F^{g_{27}} + F^{g_{57}} \,.
\ee
Note that in writing eqs.~(\ref{eq:Fu3}) and (\ref{eq:Fomu3}),
we ignored dependence on $u_3$ that might enter via $(1-u_7)$.
We could do this precisely because of the third condition in
\eqn{eq:CObranchcut}.

The three conditions~(\ref{eq:CObranchcut}) can be imposed on all 7
cyclic images of the CO surfaces, so they are effectively 21 conditions.
They are still not quite sufficient to fix all the beyond-the-symbol entries
in the coproduct tables. However, when supplemented by the branch-cut
condition~(\ref{eq:softbranchcut}) imposed on a simple soft limit
described in appendix~\ref{app:simplesoftcollinear},
and the (extended) Steinmann conditions,
they do fix all ambiguities.  Note that the space of
functions~(\ref{eq:COfunctions}) being constrained
essentially evaluate to MZVs on the constraints, and so the coefficients
of the beyond-the-symbol functions, which have MZV prefactors, are
constrained to be rational numbers, just as in the hexagon function case.

\emph{(Extended) Steinmann relations}
The heptagon Steinmann relations, as originally described at symbol
level~\cite{Dixon:2016nkn}, are simplest in the $a$-letters.
They say that a symbol first entry $a_{1i}$ should not be followed
by a second entry $a_{1j}$ with $j\in \{i+1,i+2,i+5,i+6\}$.
In ref.~\cite{Drummond:2017ssj} it was proposed that the same adjacency
conditions, natural from the cluster algebra context, should hold
for arbitrary pairs of adjacent entries, not just in the first two entries.
An alternative argument~\cite{Caron-Huot:2019bsq} for such restrictions is
based on applying the Steinmann relations on different Riemann sheets,
which differ from the original sheet by discontinuities that can be
generated at symbol level by clipping off an arbitrary number of
initial entries.  Here we want to impose the (extended) Steinmann
relations at function level.  We will impose the double coproduct constraints,
\be\label{eq:ExtStein}
F^{a_{1i},a_{1j}} = 0, \qquad j\in \{i+1,i+2,i+5,i+6\} ,
\ee
which are the function-level analogs of the symbol-adjacency constraints,
and the seven-point analog of the six-point conditions $F^{a,b}=\cdots=0$
imposed in ref.~\cite{Caron-Huot:2019bsq}.

Naively, this is not the right thing to do.  In principle,
the Steinmann relations are supposed to be imposed in kinematic regions
where two overlapping three-particle cuts are both opening up
(see e.g.~ref.~\cite{Caron-Huot:2016owq}),
while \eqn{eq:ExtStein} is being imposed everywhere.
However, in practice \eqn{eq:ExtStein} works.  Together with the integrability
and branch-cut conditions, it restricts the number of functions to
exactly the expected number: The number of extended-Steinmann
symbols at that weight, plus beyond-the-symbol functions associated with
independent MZVs multiplying lower-weight functions.
(The same is true in the six-point case~\cite{Caron-Huot:2016owq},
modulo additional complications starting at weight eight, where a few
symbols begin to drop out, due to certain MZV restrictions.)
The reason it works probably has to do with the fact that the condition
is being imposed iteratively in the weight, and also at symbol level.
Because of this, at each step \eqn{eq:ExtStein} effectively just constrains
the coefficients of weight $n-2$ constants, when imposed on weight $n$
functions. The constants are independent of the kinematics, and so the
condition can be imposed anywhere.

%%%%%%%%%%%%%%%%%%%%%%%%%%%%%%%%
\subsection{Number of beyond-the-symbol functions}
\label{sec:bsnumber}

%%%%%%%%%%%%%%%%%%%%%%%%%%%%%%%5
\renewcommand{\arraystretch}{1.25}
\begin{table}[!t]
\centering
\begin{tabular}[t]{l c c c c c c c}
\hline\hline
weight $n$
& 0 & 1 &  2 &  3 &   4 &   5 &    6 \\
\hline\hline
symbol level, parity $+$
& 1 & 7 & 28 & 91 & 280 & 791 & 2149 \\\hline
symbol level, parity $-$
& 0 & 0 &  0 &  6 &  28 & 120 &  406 \\\hline
symbol level, total
& 1 & 7 & 28 & 97 & 308 & 911 & 2555 \\\hline\hline
beyond-the-symbol, parity $+$
& 0 & 0 &  0 &  1 &   8 &  37 &  135 \\\hline
beyond-the-symbol, parity $-$
& 0 & 0 &  0 &  0 &   0 &   0 &    6 \\\hline
beyond-the-symbol, total
& 0 & 0 &  0 &  1 &   8 &  37 &  141 \\\hline\hline
\end{tabular}
\caption{Number of extended Steinmann heptagon functions through weight 6: those already seen at the symbol level, plus the beyond-the-symbol functions, graded by parity.} \label{tab:bottomup}
\end{table}
%%%%%%%%%%%%%%%%%%%%%%%%%%%%

We have carried out this procedure through weight six, leveraging the basis of Steinmann (but not extended Steinmann) symbols provided in ref.~\cite{Dixon:2016nkn}. (A basis of extended Steinmann symbols is available in ref.~\cite{Drummond:2018caf}.) The results are summarized in \Tab{tab:bottomup}, which enumerates the functions that are already visible at the symbol level, and the beyond-the-symbol functions, both graded by parity.  The symbol level numbers agree with those found previously~\cite{Drummond:2017ssj}.  In the spirit of ref.~\cite{Caron-Huot:2019bsq}, we only include independent zeta-value constants when necessary.  When they appear, they can be multiplied by lower-weight symbol-level functions to generate the tower of beyond-the symbol functions in the table.  Notice that there is no beyond-the-symbol function at weight 2, indicating that the Riemann zeta value $\z_{2}$ is not required as an independent function, but can always be absorbed into the rest of the weight 2 functions, which is exactly analogous to the hexagon function case~\cite{Caron-Huot:2019bsq}.

However, at weight 3, $\z_{3}$ \emph{does} appear as the unique beyond-the-symbol function at that weight. In the hexagon function case, $\z_{3}$ was \emph{not} an independent function, and so this difference is quite glaring. The appearance of $\z_{3}$ in the space of heptagon functions can be attributed to the weight 5 parity-odd heptagon symbols:  Once these symbols are completed to functions, the span of their $\{3,1,1\}$ double coproducts includes $\z_{3}$ (as well as every other weight 3 function).  Similarly, $\z_{4}$ is forced to be independent because it is in the span of the $\{4,1,1\}$ double coproducts of the weight 6 parity-odd functions.  It accounts for 1 of the 8 beyond-the-symbol functions at weight 4 in \Tab{tab:bottomup}; the other 7 are $\z_{3}\,\ln(u_i)$, $i=1,2,\ldots,7$.  At weight 5, the 37 beyond-the-symbol functions include $\z_{5}$ and $\z_{2}\z_{3}$, $\z_{4}\,\ln(u_i)$, and $\z_{3}$ multiplied by the weight 2 functions, for a total of $2+7+28=37$.

The first appearance of parity-odd beyond-the-symbol functions is at weight 6, where we can multiply $\z_{3}$ by the 6 weight 3 parity-odd functions.  These 6 functions are one-mass scalar hexagon integrals in six dimensions~\cite{DelDuca:2011jm}. (There are seven possible locations for the massive leg, but the cyclic sum of the seven functions vanishes.) As we will discuss further in \sect{sec:coactioncomments}, there is some evidence that these weight 6 beyond-the-symbol functions do not actually have to be independent, like $\z_{2}$ in the parity even sector.

We provide the $\{n-1,1\}$ coproducts defining the complete space of heptagon functions through weight 6 in the ancillary file \texttt{HCoproductTables.txt}.  We also need to specify the weight 8 four-loop amplitudes and their first derivatives at weight 7; the corresponding coproduct tables are contained in the files \texttt{MHE\_7.txt}, \texttt{MHO\_7.txt}, \texttt{MHE\_8.txt}, and \texttt{MHO\_8.txt}.  We specify the boundary conditions for all these functions on the CO surface in the ancillary file \texttt{HcoTable.txt}, and we present their dihedral transformations in \texttt{HDihedralSym.txt}.

%%%%%%%%%%%%%%%%%%%%%%%%%%%%%%%%%%%%%%%%%%%%%%%%%

\section{Lifting symbols of the amplitudes}
\label{sec:liftingamplitudes}

As mentioned earlier, the coefficients of the symbol-level functions in the amplitudes are known through four loops from previous work~\cite{Drummond:2014ffa,Dixon:2016nkn,Drummond:2018caf}.  To fix the coefficients of the beyond-the-symbol functions for the amplitudes, we need to impose similar constraints to those used at symbol level, but now at function level. These constraints are also easy to impose on the CO surface, or in surfaces adjacent to it, such as the soft and collinear surfaces described in appendix~\ref{app:simplesoftcollinear}.

\subsection{MHV amplitude}
\label{sec:MHVamp}

\indent\indent \emph{Soft limit.} \quad The CO surface overlaps with certain soft limits. For example, as was mentioned in \sect{sec:COsubsection}, the limit $u_4\to1$ of the CO surface is a soft limit onto the $(u,0,0)$ hexagon line. In more detail, from \eqn{eq:uisi} one can see that as the momentum $p_1\to0$, the cross ratios $u_5$ and $u_6$, which contain $s_{71}$ and $s_{12}$ in their numerators, must vanish.  In addition, it is easy to see that $u_4,u_7\to1$.  In such a limit, $u_1,u_2,u_3$ become the three cross ratios of the hexagon.  Since $u_1,u_2 \ll 1$ on the CO surface, the $u_4\to1$ limit can be identified with the soft limit with $(u,0,0)$ six-point kinematics, where the two ``0'' entries mean that the corresponding hexagon cross ratios, $v$ and $w$, are infinitesimal, while the cross ratio $u$ is generic. We identify $u=u_3$, $v=u_1$, $w=u_2$.

Alternatively, we could take the limit $p_4\to0$; then we have $u_1,u_2\to0$, $u_3,u_7\to1$, and $u_4,u_5,u_6$ become the three cross ratios of the hexagon. Again since $u_5,u_6 \ll 1$, we see that the CO surface overlaps with a second copy of the $(u,0,0)$ soft limit.

The BDS ansatz correctly reproduces the collinear and soft limits of amplitudes~\cite{Bern:2005iz}.  Therefore, in the soft limit, we expect the BDS-normalized seven-point amplitude to collapse smoothly onto the BDS-normalized six-point amplitude in the appropriate kinematics.  For MHV amplitudes, this is equivalent to saying that the seven-point remainder function goes smoothly to the six-point remainder function in the limit:
\be\label{eq:R7softlimit}
R_7^{(L)}\Big|_{{\rm CO},\ u_4\to1}\ \to\ R_6^{(L)}(u_1,u_2,u_3)\Big|_{u_{1,2} \ll 1} \,.
\ee
We can then use the known results from the hexagon function bootstrap to constrain the seven-point amplitude.

\emph{$\bar{Q}$ final entries.} \quad The final-entry condition arises from certain anomaly equations (the $\bar{Q}$ equations) for dual superconformal symmetry generators~\cite{Bullimore:2011kg,CaronHuot:2011kk}.  The MHV final-entry condition, which was used to bootstrap the three- and four-loop MHV symbols~\cite{Drummond:2014ffa,Dixon:2016nkn}, states that only the 14 final entries $a_{2j}$ and $a_{3j}$ are allowed.

For the MHV case, dihedral invariance, the final-entry condition, and the $(u,0,0)$ soft limit turn out to be all we need to fix the beyond-the-symbol terms in the amplitude through four loops.  Note that the four-loop beyond-the-symbol ambiguities proportional to $\z_6$ or $(\z_3)^2$ ($\z_4$) are weight 2 (4) symbols, which were also encountered at one loop (two loops) in ref.~\cite{Dixon:2016nkn} and are known to be highly constrained even before taking the soft limit (see Table 3 there): there is only a single parameter left at weight 2 and at weight 4.  At odd weights, 1, 3 and 5, we find that there are \emph{no} free parameters left after imposing dihedral invariance and the final-entry condition!  So the only job of the $(u,0,0)$ soft limit at four loops is to fix 3 parameters, the coefficients of the one and two loop expressions multiplied by the appropriate zeta values, plus the 4 weight 8 constant MZVs.  (The one and two loop expressions are singular in the soft limits because they represent BDS-like normalized amplitudes.)  In practice, the four loop determination is more complicated because our basis only extends to weight 6, so we first fix the $\{6,1,1\}$ coproducts and then integrate up from there, along the lines of how the seven-loop six-point MHV amplitude was constructed~\cite{Caron-Huot:2019vjl}.

We have also checked that the MHV amplitudes behave correctly in the alternative soft limit~(\ref{eq:soft1vvgletters}) where the six-point cross ratios become $(u,v,w)=(1,u_2,u_2)$, and in the collinear limit~(\ref{eq:collgletters}).

For example, at two loops, the six-point $(u,0,0)$ soft target
can be written in terms of classical polylogarithms~\cite{Goncharov:2010jf},
\be\label{eq:R62u00}
\begin{split}
R_6^{(2)}( u_{1,2}&\ll 1, u_3 ) = \\[1ex]
& 6 \, {\rm Li}_4\left(\frac{-u_3}{1-u_3}\right)
- \frac{1}{2} \, \left[ {\rm Li}_2\left(\frac{-u_3}{1-u_3}\right) \right]^2
- 2 \, \ln\left(\frac{u_3}{u_1 u_2}\right)
   \ {\rm Li}_3\left(\frac{-u_3}{1-u_3}\right)
\\[1ex]
&+ \Biggl[ \ln\left(\frac{1-u_3}{u_1 u_2}\right) \ln u_3
            + \ln u_1 \ln u_2 - \frac{1}{2} \, \ln^2(1-u_3)
            - 3 \, \zeta_2 \Biggr]
  \, {\rm Li}_2\left(\frac{-u_3}{1-u_3}\right)
\\[1ex]
&+ \frac{1}{8} \ln^4(1-u_3)
- \Bigl[ \frac{1}{6} \, \ln^2(1-u_3) + \zeta_2 \Bigr]
 \ln(1-u_3) \, \Bigl[ \ln u_3 + 2 \, \ln(u_1 u_2) \Bigr]
\\[1ex]
&+ \frac{1}{2} \ln^2(1-u_3)
   \, \Bigl[ \ln u_1 \, \ln u_2 + \ln(u_1 u_2) \, \ln u_3 + 3 \, \zeta_2 \Bigr]
\\[1ex]
&- \ln(1-u_3) \, \ln u_3 \, \ln u_2 \, \ln u_1
\\[1ex]
&+ \frac{\zeta_2}{2}
   \Bigl[ 2 \, \ln(u_1 u_2) \, \ln u_3
        + 2 \, \ln u_1 \, \ln u_2 \Bigr]
+ \frac{17}{4} \zeta_4 \,.
\end{split}
\ee

The two-loop seven-point remainder function on the CO surface can
also be written in terms of classical polylogarithms:
\be\label{eq:R72CO}
\begin{split}
R_7^{(2)}( u_{1,2,5,6}&\ll 1, \, u_7=1 ) = \\[1ex]
& 6 \, {\rm Li}_4\left(\frac{-u_3}{1-u_3}\right)
- \frac{1}{2} \, \left[ {\rm Li}_2\left(\frac{-u_3}{1-u_3}\right) \right]^2
- 2 \, \ln\left(\frac{u_3}{u_1 u_2 u_4}\right)
   \ {\rm Li}_3\left(\frac{-u_3}{1-u_3}\right)
\\[1ex]
&+ \Biggl[ \ln\left(\frac{1-u_3}{u_1 u_2 u_4}\right) \ln u_3
            + \ln(u_1 u_4) \ln u_2 - \frac{1}{2} \, \ln^2(1-u_3)
            - 3 \, \zeta_2 \Biggr]
  \, {\rm Li}_2\left(\frac{-u_3}{1-u_3}\right)
\\[1ex]
&+ \frac{1}{8} \ln^4(1-u_3)
- \Bigl[ \frac{1}{6} \, \ln^2(1-u_3) + \zeta_2 \Bigr]
 \ln(1-u_3) \, \Bigl[ \ln u_3 + 2 \, \ln(u_1 u_2 u_4) \Bigr]
\\[1ex]
&+ \frac{1}{2} \ln^2(1-u_3)
   \, \Bigl[ \ln u_1 \, \ln u_2 + \ln(u_1 u_2) \, \ln u_3
      + \ln(u_2 u_3) \, \ln u_4 + 3 \, \zeta_2 \Bigr]
\\[1ex]
&- \ln(1-u_3) \, \ln u_3 \, \ln u_2 \, \ln(u_1 u_4)
\\[1ex]
&+ \frac{\zeta_2}{2}
   \Bigl[ \Bigl( \ln u_4 + 2 \, \ln(u_1 u_2 u_5) \Bigr) \ln u_3
        + 2 \, \ln u_1 \, \ln u_2 \Bigr]
+ \frac{17}{4} \zeta_4
\\[2ex]
&+ \{\ u_1 \leftrightarrow u_6,\ u_2 \leftrightarrow u_5,
    \ u_3 \leftrightarrow u_4\ \} .
\end{split}
\ee
It is not hard to verify that as $u_4 \to 1$ this expression collapses
to \eqn{eq:R62u00}.
The three- and four-loop results, $R_7^{(3)}$ and $R_7^{(4)}$ on the CO surface,
can be found in our ancillary file \texttt{R\_P\_o\_co.txt}.

We can now compare \eqn{eq:R72CO} to the previously known two-loop MHV amplitude~\cite{Golden:2014xqf} in generic kinematics.  That formula contains a large number of terms, most of which are classical, but 112 of which feature the following non-classical polylogarithmic function,
\be
L_{2,2}(x,y) \ \equiv\ \frac{1}{2} \int_0^1 \frac{dt}{t}
 \Bigl[ {\rm Li}_2(-tx) \, {\rm Li}_1(-ty)
      - {\rm Li}_2(-ty) \, {\rm Li}_1(-tx) \Bigr] \,.
\ee
When we drop onto the CO surface, the 112 instances of $L_{2,2}$ reduce to just three pairs, and the non-classical parts of $L_{2,2}$ cancel pairwise by using the identity
\be
\begin{split}
L_{2,2}(x,y)\ +\ & L_{2,2}(1/y,1/x) = \\[1ex]
& 3 \, {\rm Li}_4(x/y) + {\rm Li}_4(-x) - {\rm Li}_4(-y)
\\[1ex]
& - \ln(x/y) \, {\rm Li}_3(x/y) - \ln y \, {\rm Li}_3(-x)
 + \ln x \, {\rm Li}_3(-y)
\\[1ex]
& + \frac{1}{4} \, ( \ln^2 y + 2 \zeta_2 ) \, {\rm Li}_2(-x)
 - \frac{1}{4} \, ( \ln^2 x + 2 \zeta_2 ) \, {\rm Li}_2(-y)
\\[1ex]
& + \frac{1}{24} \, \ln^2 y \, \ln(x/y) \, ( \ln y - 3 \, \ln x )
 - \frac{\zeta_2}{4} \, \ln(x/y) \, ( \ln x - 3 \, \ln y )
 - 3 \, \zeta_4 \,.
\end{split}
\ee
Using also standard identities for classical polylogarithms, we obtain
complete agreement with our result \eqn{eq:R72CO}.

%%%%%%%%%%%%%%%%%%%%%%%%%
\subsection{NMHV amplitude}
\label{sec:NMHVamp}

We build an ansatz for the components $E_0$, $E_{12}$ and $E_{14}$, on which we impose a number of different constraints:

\indent\indent \emph{1. Dihedral symmetry.} \quad The component $E_0$ has full dihedral symmetry, while $E_{12}$ and $E_{14}$ are each only invariant under a flip (the ones fixing $u_6$ and $u_7$, respectively).  Cyclic permutations of $E_{12}$ and $E_{14}$ sweep out the remaining 12 components $E_{i,i+1}$ and $E_{i,i+3}$.

\indent\indent \emph{2. Soft limit.} \quad  To analyze the soft limits of the NMHV amplitude, we also need to take soft limits of the $R$-invariants. In practice, this is easier to do in the special case of \emph{soft-collinear} limits. In terms of momentum twistors, this means that for the $p_1\to0$ limit we take either $\cZ_1\sim\cZ_7$ or $\cZ_1\sim\cZ_2$. (See appendix \ref{app:softcollinear} for a detailed description of the general $p_1\to0$ soft limit.) The two soft-collinear limits are equivalent, as it must be in order for the BDS-normalized amplitude to have a smooth soft limit. If we take $\cZ_1\sim\cZ_2$, then the $R$-invariants either vanish or reduce to those with only \{2,3,4,5,6,7\} in the entries; therefore they become six-particle $R$-invariants. In this limit we will use the notation
\be
(7) \equiv [23456] \,, \quad\quad \text{etc.}
\ee
Using the six-term identity,
\be
(2) - (3) + (4) - (5) + (6) - (7) = 0
\ee
to eliminate the six-particle $R$-invariant (6), and recalling that the soft limit corresponds to $u_4,u_7 \to 1$ and $u_5,u_6 \ll 1$ in the cross ratios, we see that in the soft limit,
\be
\begin{split} \label{eq:BSoftLimit}
\left.B\right|_{p_1\to 0} &= \biggl.\biggl\{
[B_{12} - B_{62}] \cdot (2) + [B_0 + B_{23} + B_{62}] \cdot (3)
+ [B_{14} - B_{62}] \cdot (4) 
\\
& \quad
  + [B_0 + B_{51} + B_{25} + B_{62}] \cdot (5)
  + [B_0 + B_{71} + B_{62}] \cdot (7) \biggr\}\biggr|_{u_4,u_7\to 1,\ u_5,u_6\ll 1}\,.
\end{split}
\ee
Here $B$ denotes the BDS-normalized NMHV amplitude~(\ref{eq:B7components}). Additionally, we require $u_1,u_2\ll1$ in order for the soft limit to intersect the CO surface.  We can now fit the coefficients of the six-particle $R$-invariants to the known hexagon NMHV components, and use them as constraints on the heptagon components.

To be specific, the six-point BDS-normalized NMHV amplitude can be decomposed as
\be
\begin{split}
B^{\text{6-point}}(u,v,w) &= \frac{1}{2}\biggl[ B'(u,v,w)\bigl[(2) + (5)\bigr] + B'(v,w,u)\bigl[(3) + (6)\bigr] + B'(w,u,v)\bigl[(4) + (7)\bigr] \\
& \qquad
+ \widetilde{B}(u,v,w)\bigl[(2) - (5)\bigr] - \widetilde{B}(v,w,u)\bigl[(3) - (6)\bigr] + \widetilde{B}(w,u,v)\bigl[(4) - (7)\bigr] \biggr]\,,
\end{split}
\ee
in the same fashion as in refs.~\cite{Dixon:2015iva,Caron-Huot:2019vjl}, where $u,v,w$ are the three cross ratios for six particles, and $B'$ and $\widetilde{B}$ are parity-even and parity-odd functions, respectively. As in the MHV case, we identify $u=u_3$, $v=u_1$, $w=u_2$. In practice, we match to the six-point amplitude in the special kinematics $u_3 \equiv u\to1$ and $u_1 = u_2 = v = w \to 0$. In these kinematics, the parity odd function $\widetilde{B}$ vanishes, and $B'(u,v,w)$ satisfies $B'(u_2,u_2,1) = B'(1,u_2,u_2) = -B'(u_2,1,u_2)$, due to symmetry and collinear behavior~\cite{Dixon:2014iba,Dixon:2015iva}. Again using the six-term identity to relate $R$-invariants, the six-point amplitude simplifies to
\be
B^{\text{6-point}}\ =\ B'(1,u_2,u_2)|_{u_2\to0} \, \bigl[(3) - (4) + (5)\bigr] \,.
\ee
Matching this expression to the seven-point soft limit (\ref{eq:BSoftLimit}) on the CO surface, with $u_3 \to 1$, $u_1=u_2 \ll 1$, then generates the following constraints for the seven-particle NMHV components:
\bea
B_{12} - B_{62} &\to& 0, \label{eq:B12soft}\\
B_0 + B_{23} + B_{62} &\to& B'(1,u_2,u_2)|_{u_2\to0} \,, \label{eq:B23soft}\\
B_{14} - B_{62} &\to& -B'(1,u_2,u_2)|_{u_2\to0} \,, \label{eq:B14soft}\\
B_0 + B_{51} + B_{25} + B_{62} &\to& B'(1,u_2,u_2)|_{u_2\to0} \,,
\label{eq:B51soft}\\
B_0 + B_{71} + B_{62} &\to& 0. \label{eq:B71soft}
\eea
Because of the $u_2\to0$ limit, the right-hand side contains only logarithms of $u_2$ and zeta values. The targets at 2, 3 and 4 loops are:
\bea
B^{\prime\,(2)}(1,u_2,u_2)|_{u_2\to0} &=&
\frac{1}{4} \ln^4 u_2 - 3 \z_2 \ln^2 u_2
+ 4 \z_3 \ln u_2 + \frac{5}{2} \z_4 \,, \label{eq:Bprime2}\\
B^{\prime\,(3)}(1,u_2,u_2)|_{u_2\to0} &=&
\frac{1}{36} \ln^6 u_2 - \frac{5}{3} \z_2 \ln^4 u_2 + \frac{71}{2} \z_4 \ln^2 u_2
- 16 (2 \z_5+\z_2 \z_3) \ln u_2
\nonumber\\&&\hskip0.2cm\null
- \frac{77}{2} \z_6 + 4 (\z_3)^2 
\,, \label{eq:Bprime3}\\
B^{\prime\,(4)}(1,u_2,u_2)|_{u_2\to0} &=&
\frac{1}{576} \ln^8 u_2 - \frac{7}{24} \z_2 \ln^6 u_2
- \frac{1}{3} \z_3 \ln^5 u_2 + \frac{1285}{48} \z_4 \ln^4 u_2
\nonumber\\&&\hskip0.2cm\null
+ \frac{1}{3} ( 14 \z_5 - 16 \z_2 \z_3 ) \ln^3 u_2
- \frac{1}{16} ( 5935 \z_6 + 160 (\z_3)^2 ) \ln^2 u_2
\nonumber\\&&\hskip0.2cm\null
+ (300 \z_7 + 156 \z_2 \z_5 + 174 \z_4 \z_3) \ln u_2
\nonumber\\&&\hskip0.2cm\null
+ \frac{48935}{96} \z_8 - 64 \z_3 \z_5 - 4 \z_2 (\z_3)^2 
\,. \label{eq:Bprime4}
\eea

\emph{3. No spurious poles.} \quad  We require that the only physical singularities in the NMHV amplitude correspond to $x_{i,j}^2 \to 0$.  In terms of momentum twistors, these locations are where four-brackets of the form $\ket{i-1,i,j-1,j}$ vanish.  On the other hand, in the BDS-like-normalized amplitude~(\ref{eq:E7components}) the $R$-invariants~$(ij)$, defined in \eqn{fivebrak}, contain four-brackets in their denominators which are not of the form above.  We require that their residues cancel in the amplitude. The resulting constraints on the components $E_{ij}$ were already employed at symbol level~\cite{Dixon:2016nkn,Drummond:2018caf}; we simply record them here:
\be
\begin{split}\label{eq:spuriousbulk}
&\text{Spurious I:} \quad   E_{47}\vert_{\ket{1356} =0} = 0\,, \\
&\text{Spurious II:} \quad  (E_{23} - E_{25})\vert_{\ket{1467}=0} = 0\,,
\end{split}
\ee
plus all cyclic images of these relations.

To impose these conditions on the CO surface, we need to identify where it intersects the various surfaces of vanishing 4-brackets.  For example, using eqs.~(\ref{aletters}), (\ref{gletters}) and (\ref{eq:glettersu}), we see that
\bea
1-u_3 &=& g_{23} = \frac{\ket{1467}\ket{3457}}{\ket{6734}\ket{4571}} \,,
\label{eq:g23}\\
1-u_4 &=& g_{24} = \frac{\ket{1456}\ket{7125}}{\ket{5612}\ket{4571}} \,.
\label{eq:g24}
\eea
Thus $\ket{1467}=0$ and $\ket{3457}=0$ both correspond to the line $u_3=1$ when restricted to the CO surface, whereas $\ket{1456}=0$ and $\ket{7125}=0$ correspond to $u_4=1$. Similarly,
\bea
1-u_2u_5-u_4u_7 &=& g_{41} = \frac{a_{47}a_{57}}{a_{12}a_{15}a_{17}}
= \frac{\ket{6134}\ket{7(12)(34)(56)}}{\ket{6734}\ket{7134}\ket{1256}} \,,
\label{eq:g41}\\
1-u_4u_7-u_6u_2 &=& g_{43} = \frac{a_{42}a_{52}}{a_{14}a_{17}a_{12}}
= \frac{\ket{1356}\ket{2(34)(56)(71)}}{\ket{1256}\ket{2356}\ket{3471}} \,.
\label{eq:g43}
\eea
The left-hand sides both become $1-u_4$ as we approach the CO surface; hence
$\ket{6134}=0$ and $\ket{1356}=0$ both intersect the CO surface at $u_4=1$.
Finally, from \eqn{eq:COgletters} one can show that on the CO surface
\be
1-u_4 \approx \sqrt{\frac{g_{27}g_{15}}{g_{57}g_{26}g_{56}g_{11}}}
= \frac{a_{32}a_{14}}{a_{22}a_{17}}
= \frac{\ket{1345}\ket{6712}\ket{2356}}{\ket{2345}\ket{1256}\ket{1367}} \,,
\label{eq:g2757etc}
\ee
and therefore $\ket{1345}=0$ also corresponds to $u_4=1$.

Combining these kinematic relations, and their CO-preserving flips, with the cyclic images of the bulk spurious pole conditions~(\ref{eq:spuriousbulk}), we obtain the following constraints on the CO surface:
\bea
E_{47} &=& E_{25} = E_{34} - E_{36} = E_{67} - E_{62} = E_{23} - E_{73} = 0
\quad\text{for\ }u_4=1, \label{eq:spuriousCOA}\\
E_{51} &=& E_{73} = E_{12} - E_{62} = E_{56} - E_{36} = E_{23} - E_{25} = 0
\quad\text{for\ }u_3=1, \label{eq:spuriousCOB}
\eea
where the second set of equations is related to the first by the flip~(\ref{eq:COflip}).

\emph{4. $\bar{Q}$ final entries.} \quad In the NMHV case, the constraints arising from the $\bar{Q}$ equations~\cite{CaronHuot:2011kk} allow for 147 distinct ($R$-invariant) $\times$ (final entry) combinations in $E$, which are recorded in ref.~\cite{Dixon:2016nkn}. Again we require these conditions to also be satisfied for beyond-the-symbol functions.

Taken together, the above constraints allow us to fix the NMHV amplitude through three loops.  At four loops we also employed another soft limit and a collinear limit, described in appendix~\ref{app:simplesoftcollinear}, to provide additional matching constraints.  We also used the OPE for the scalar $(7145)$ component~\cite{Basso:2013aha,Bassoprivate} to fix the coefficients of two surviving, purely beyond-the-symbol ambiguity functions at four loops. In the ancillary file \texttt{R\_P\_o\_co.txt}, we provide the values of the components of the NMHV amplitude --- or rather the ratio function --- on the CO surface through four loops.  The file \texttt{AmpsH.txt} gives the MHV and NMHV amplitudes in terms of our basis of heptagon functions.

%%%%%%%%%%%%%%%%%%%%%%%%%%%%%%%%%%%

\section{Into the bulk}
\label{sec:bulk}

Now that we have fixed the full function-level coproduct table, as well as specified the amplitudes on a boundary (the CO surface), we can integrate along any line extending from the CO surface to obtain values of the amplitudes in the bulk, i.e.~where all cross ratios are finite and the BDS- or BDS-like normalized amplitudes are finite.  In this paper we confine our numerical studies to the Euclidean bulk region.

%%%%%%%%%%%%%%%%%%%%%
\subsection{The diagonal line from the origin}

First, let us try to make the kinematics as symmetric as possible,
given the Gram determinant constraint~(\ref{eq:COGramdet}).
We let $u_1=u_2=\dots=u_6=u$, and solve the Gram determinant constraint
for $u_7$,
\be\label{eq:u7symline}
u_7 = \frac{(1-u-u^2)^2}{1-2 u^2}.
\ee
We call this the diagonal (or ``symmetric'') line.
It intersects the origin, a part of the CO surface, for $u \to 0$
and $u_7 \to 1$.
For finite $u$, the symbol letters become complicated: The even letters contain 
fifth-order polynomials in $u$ and the odd letters contain
the square root of a product of cubic polynomials in $u$.

\begin{figure}[h]
\center
\includegraphics[width=0.95\linewidth]{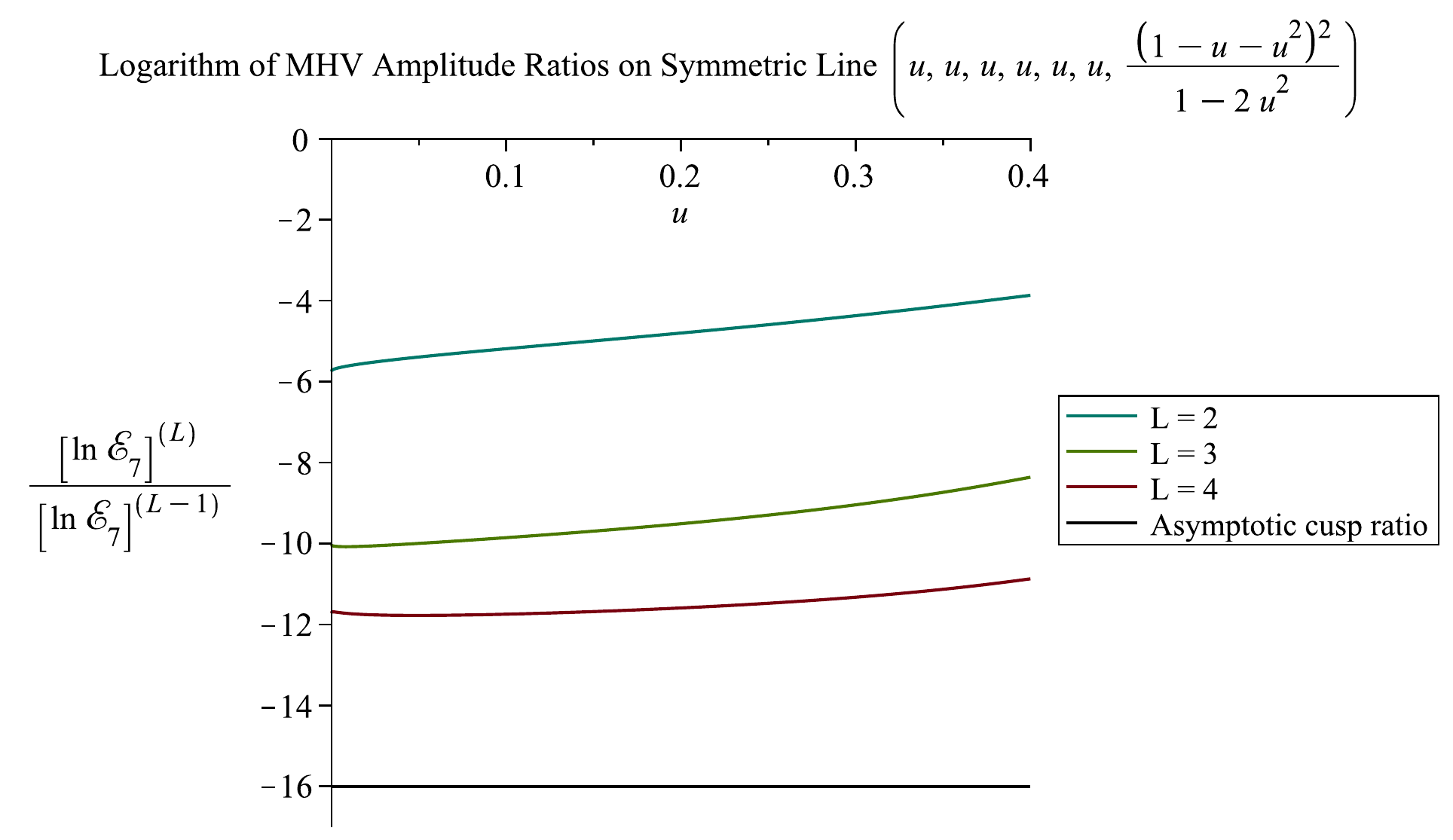}
\caption{Ratios of logarithms of the BDS-like normalized MHV amplitude
${\cal E}_7$ at successive loop orders, on the diagonal line where $u_1=\dots=u_6=u$.}
\label{Fig:lnEMHVsymline}
\end{figure}

However, since we are interested in numerical values on a one-dimensional line, it is straightforward to do a power series expansion around $u=0$.  We use the coproduct tables and the chain rule to write the $u$ derivatives of all functions in the basis in terms of the functions at one lower weight.  We integrate up term-by-term in the expansion, and weight-by-weight, fixing the boundary conditions at the origin. An expansion with forty terms suffices for convergence past the boundary of the positive region at $u \approx 0.35689$, where all seven cross ratios become equal. At two loops, our series expansion is in excellent numerical agreement with the results of ref.~\cite{Golden:2014xqf}.  In \Fig{Fig:lnEMHVsymline} we plot on this line the ratio of logarithms of the MHV amplitude at successive loop orders, $[\ln \cE_7]^{(L)}/[\ln \cE_7]^{(L-1)}$. The ratio depends quite weakly on $u$. As the loop order increases, it trends in the direction of the asymptotic ratio of successive terms for the cusp anomalous dimension,
\be\label{eq:asymcuspratio}
\frac{\Gcusp^{(L)}}{\Gcusp^{(L-1)}} \to -16, \qquad \text{as $L\to\infty$.}
\ee
This trending behavior is typical of six-point amplitudes away from boundaries, where results are available through seven loops~\cite{Caron-Huot:2019vjl}.  Now we can start to see it at seven points for the first time.

\begin{figure}[h]
\center
\includegraphics[width=0.7\linewidth]{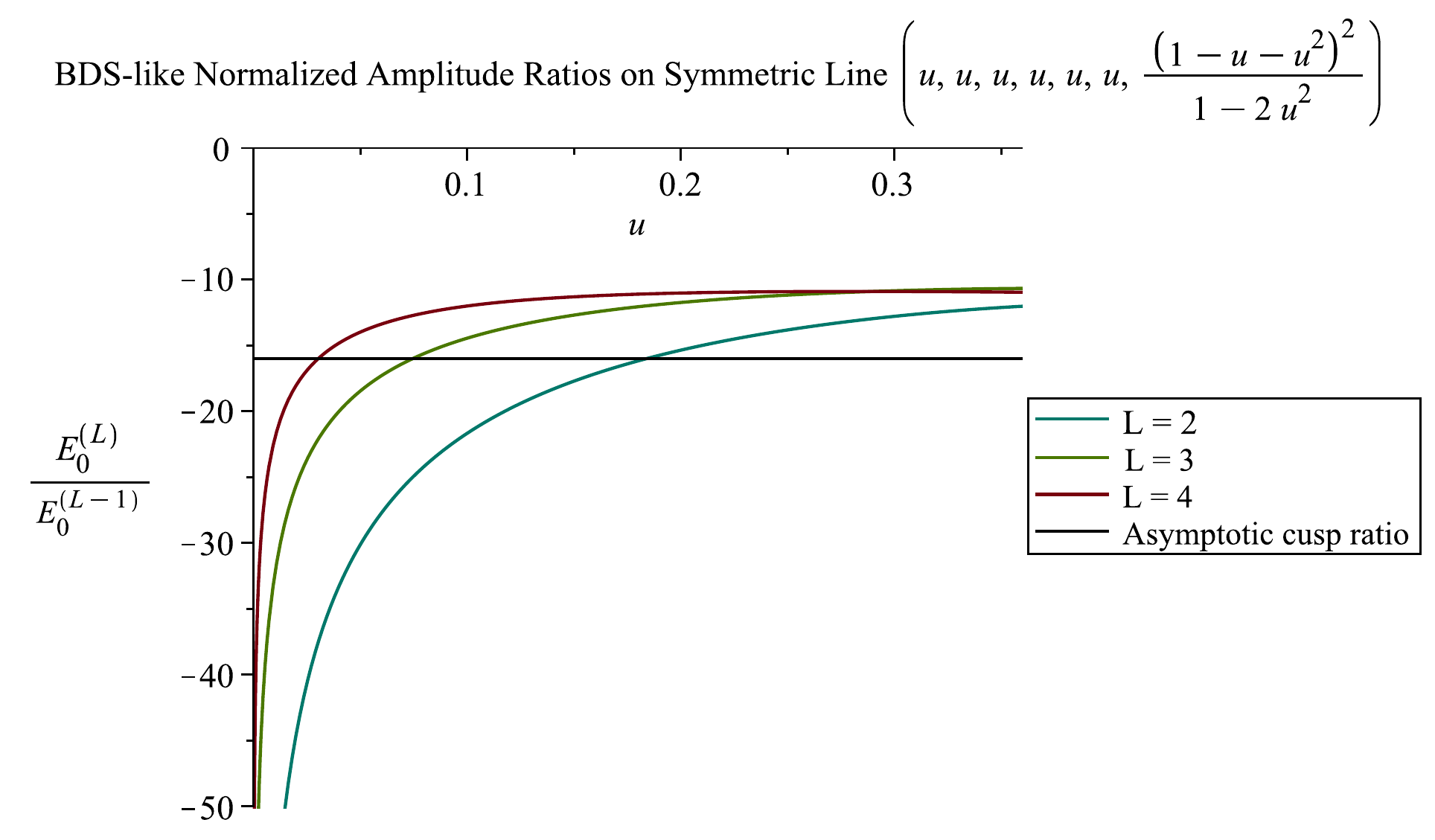}
\includegraphics[width=0.495\linewidth]{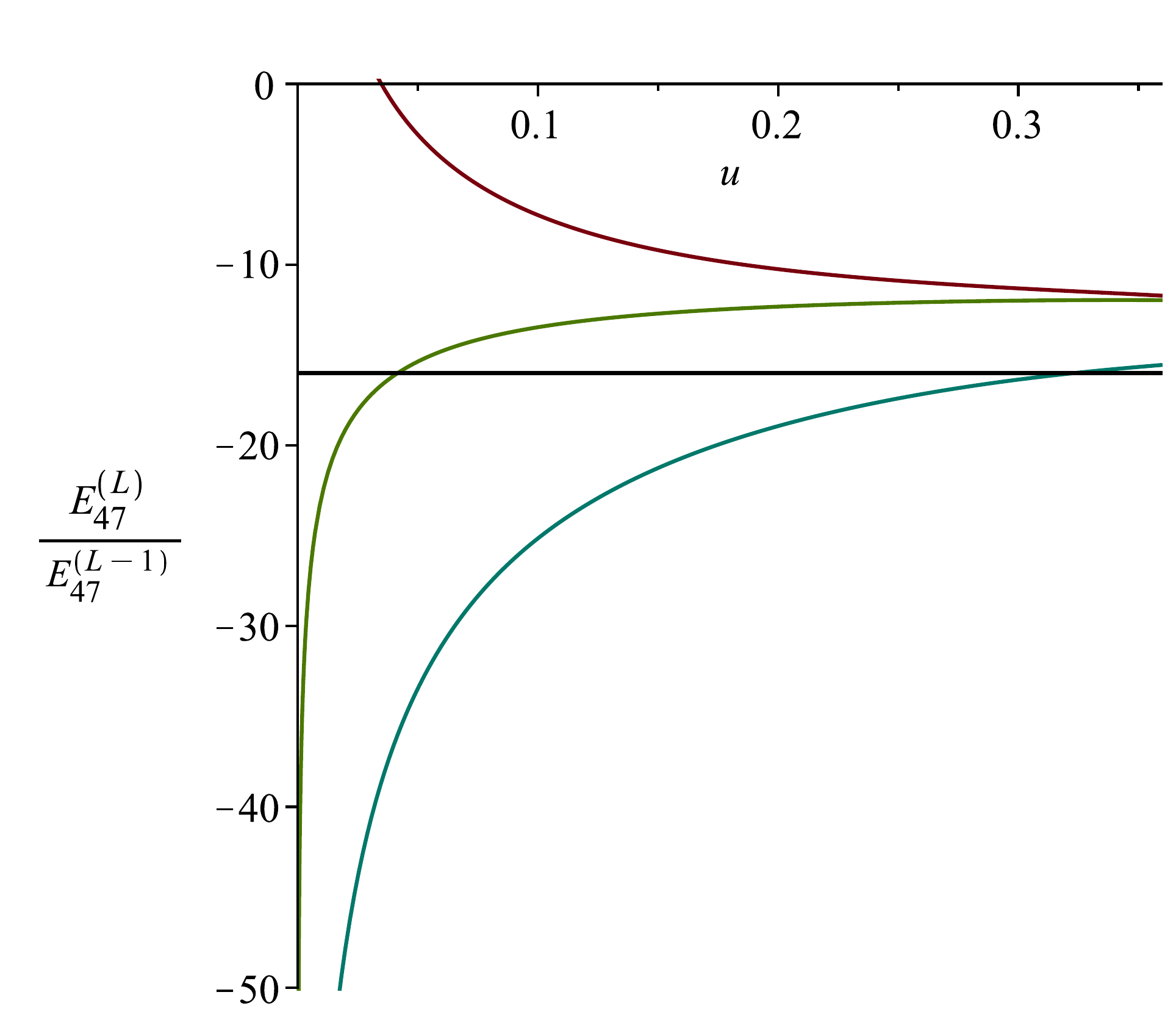}
\includegraphics[width=0.495\linewidth]{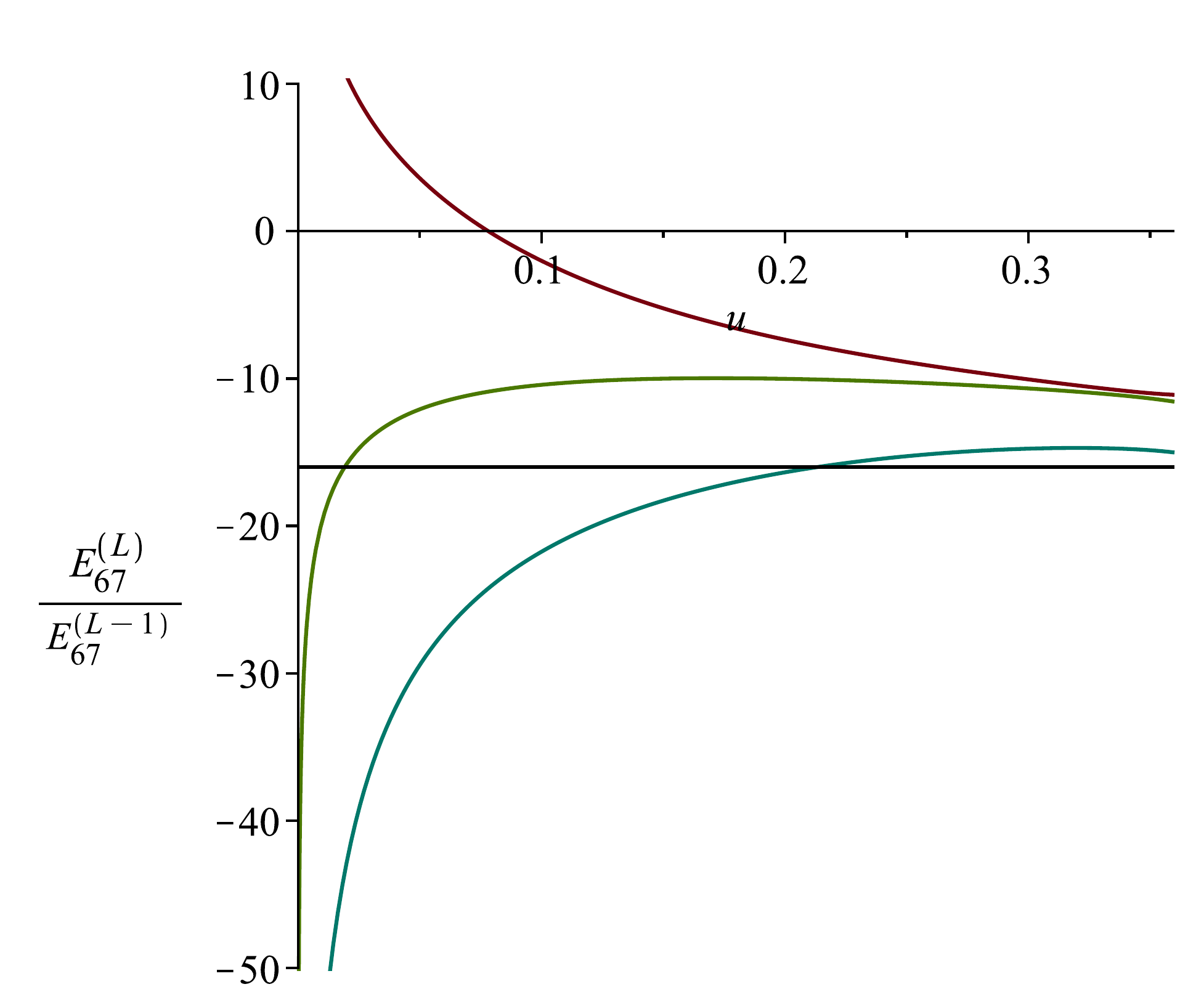}
\caption{Ratios of the BDS-like normalized NMHV amplitude components $E_0$, $E_{47}$ and $E_{67}$ at successive loop orders, on the diagonal line $u_1=\dots=u_6=u$.}
\label{Fig:NMHVsymline}
\end{figure}

In \Fig{Fig:NMHVsymline}, we plot the ratios of three different BDS-like-normalized NMHV components at successive loop orders.  In this case the logarithmic divergences at the origin, $u\to0$ are different at each order (see \sect{sec:origin}), and so one has to go to larger values of $u$ before the curves flatten.  Also, the trend toward the asymptotic cusp ratio value of $-16$ is not as clear as it is for $\ln \cE_7$.

%%%%%%%%%%%%%%%%%%%%%%%%%%%%%%%

\subsection{Self-crossing lines}

Next we evaluate the amplitudes on two lines contained within the self-crossing surface (on the Euclidean sheet).  On a physical sheet, the surface corresponds to light-like polygonal Wilson loops in the dual picture that develop a self-crossing, and there would be a singularity as the configuration is approached. In the Euclidean version, though, the amplitudes remain finite. For the heptagon case, the self-crossing surface is four-dimensional, and it can be parametrized by $u_1,u_2,u_5,u_6$; the other cross ratios are related by \cite{Dixon:2016epj},
\be
\text{\bf{Self-crossing:}} \qquad
u_3 = 1 - \frac{u_1u_5}{u_6},\qquad
u_4 = 1 - \frac{u_2u_6}{u_1},\qquad
u_7 = 1\,.
\ee
The self-crossing surface intersects the CO surface at $u_3 \to 1, u_4 \to 1$.

The first line we consider in this surface is
\begin{align}
\label{eq:line1}
\text{\bf{Line I:}} \quad
\begin{cases}
u_1,u_2,u_5,u_6 &= u, \\
u_3,u_4 &= 1-u, \\
u_7 &= 1.
\end{cases}
\end{align}
It intersects the CO surface as $u\to0$. Moreover, the other end of the line, $u\to1$ corresponds to a soft limit onto six-point kinematics where all three cross ratios go to unity, $(u,v,w)=(1,1,1)$.

The symbol alphabet on the above line contains square roots ($\Delta\ne 0$), which make it difficult to find explicit representations of the functions. However, we know the behavior on both ends of the line, and we can integrate up using power series expansions in $u$ and $1-u$, respectively, from each end.

Here is a second example of a line on the self-crossing surface,
\begin{align}
\label{eq:line2}
\text{\bf{Line II:}} \quad
\begin{cases}
u_1,u_6 &= \frac{4u}{(1+u)^2}, \\
u_2,u_5 &= u, \\
u_3,u_4 &= 1-u, \\
u_7 &= 1.
\end{cases}
\end{align}
It has the same endpoints as Line I at $u=0$ and $u=1$. Furthermore, the symbol alphabet rationalizes over $u$, and the parity odd letters all become trivial, $g_{5i} = g_{6i} = 1$, on the line ($\Delta=0$). We have been able to find a two-dimensional surface containing this line on which the symbol alphabet becomes \emph{linearly reducible} \cite{Brown:2008um}. It is then possible to find explicit $G$-function representations of the amplitudes on the surface, using a \emph{fibration basis} constructed algorithmically~\cite{Panzer:2015ida} (see also references in ref.~\cite{Duhr:2019tlz}). The procedure can be performed with computer programs such as \textsc{PolyLogTools} \cite{Duhr:2019tlz}.

\begin{figure}[h]
\center
\includegraphics[width=0.75\linewidth]{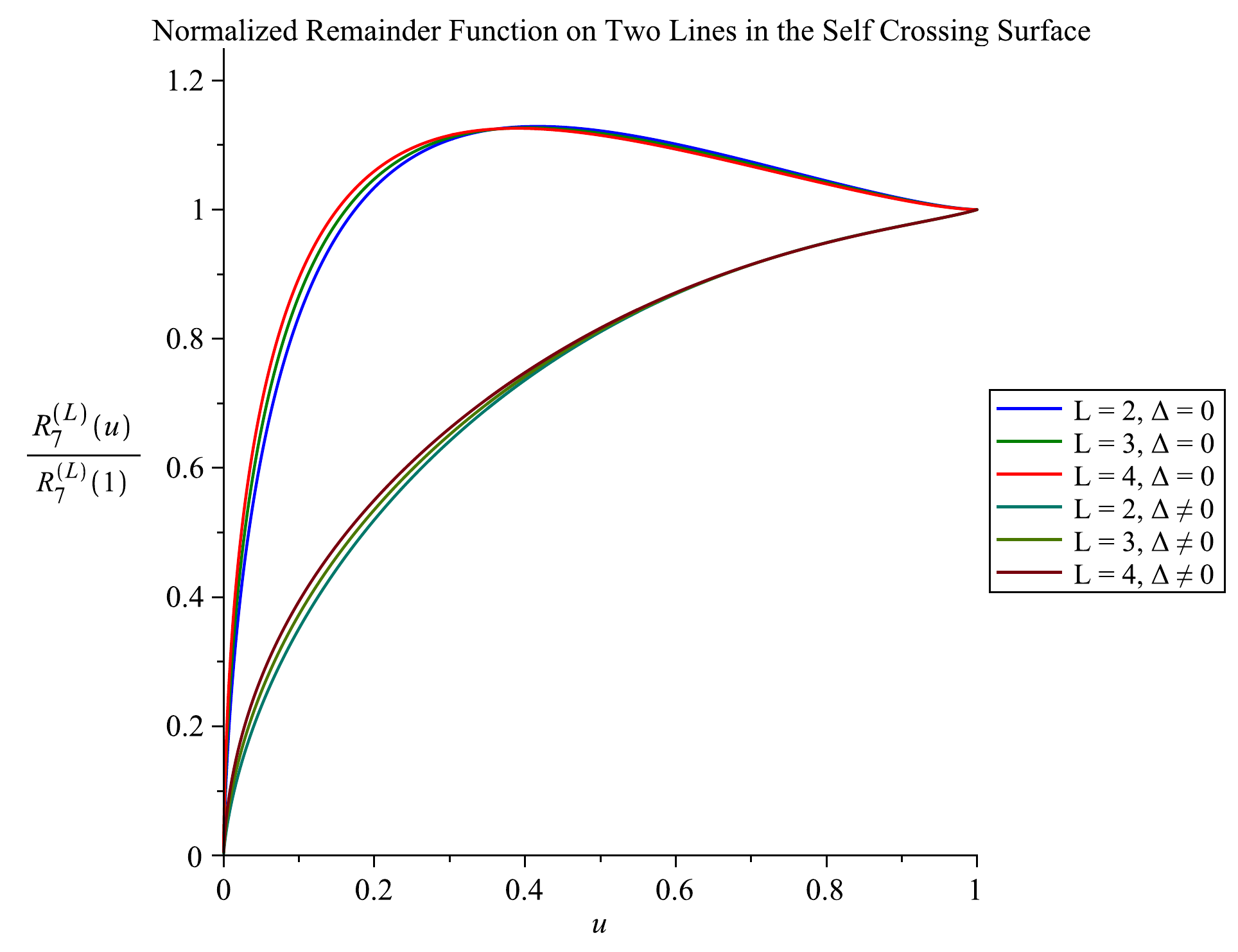}
\caption{The $L$-loop remainder function $R_7^{(L)}$ on Line I ($\Delta\ne0$) and on Line II ($\Delta=0$), normalized by its value at $u=1$.}
\label{Fig:R7Lselfcross}
\end{figure}

In \Fig{Fig:R7Lselfcross}, we plot the remainder function $R_7^{(L)}$ at loop orders $L=2,3,4$ on Lines I and II, after normalizing by its value at $u=1$ (which is equal to $R_6^{(L)}(1,1,1)$). The remainder function vanishes at $u=0$ because it is a double soft limit.  We see that the normalized remainder function has quite similar shapes at different loop orders on each line, but it behaves differently on the two lines.

\begin{figure}[ht]
\center
\includegraphics[width=0.7\linewidth]{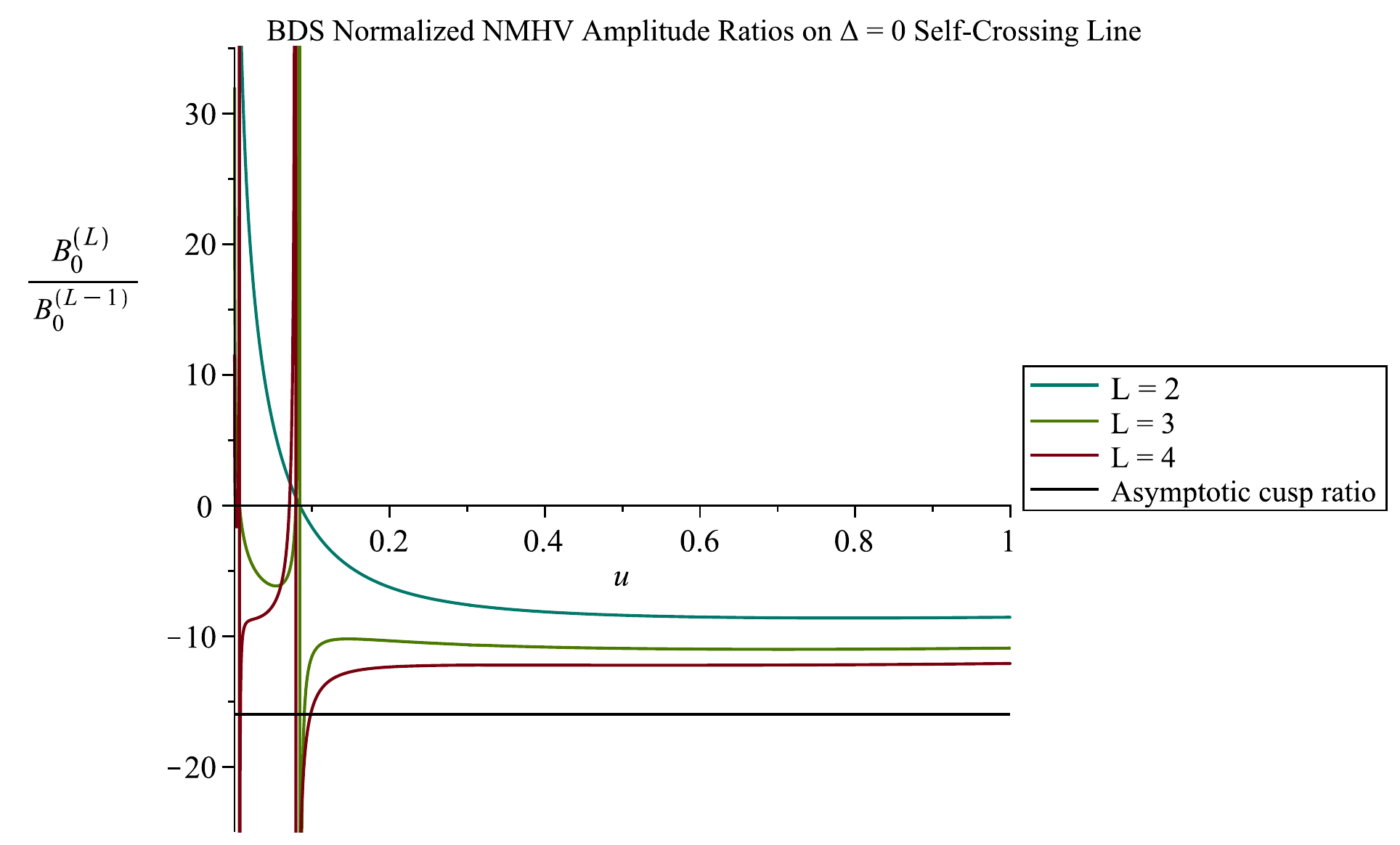}
\includegraphics[width=0.495\linewidth]{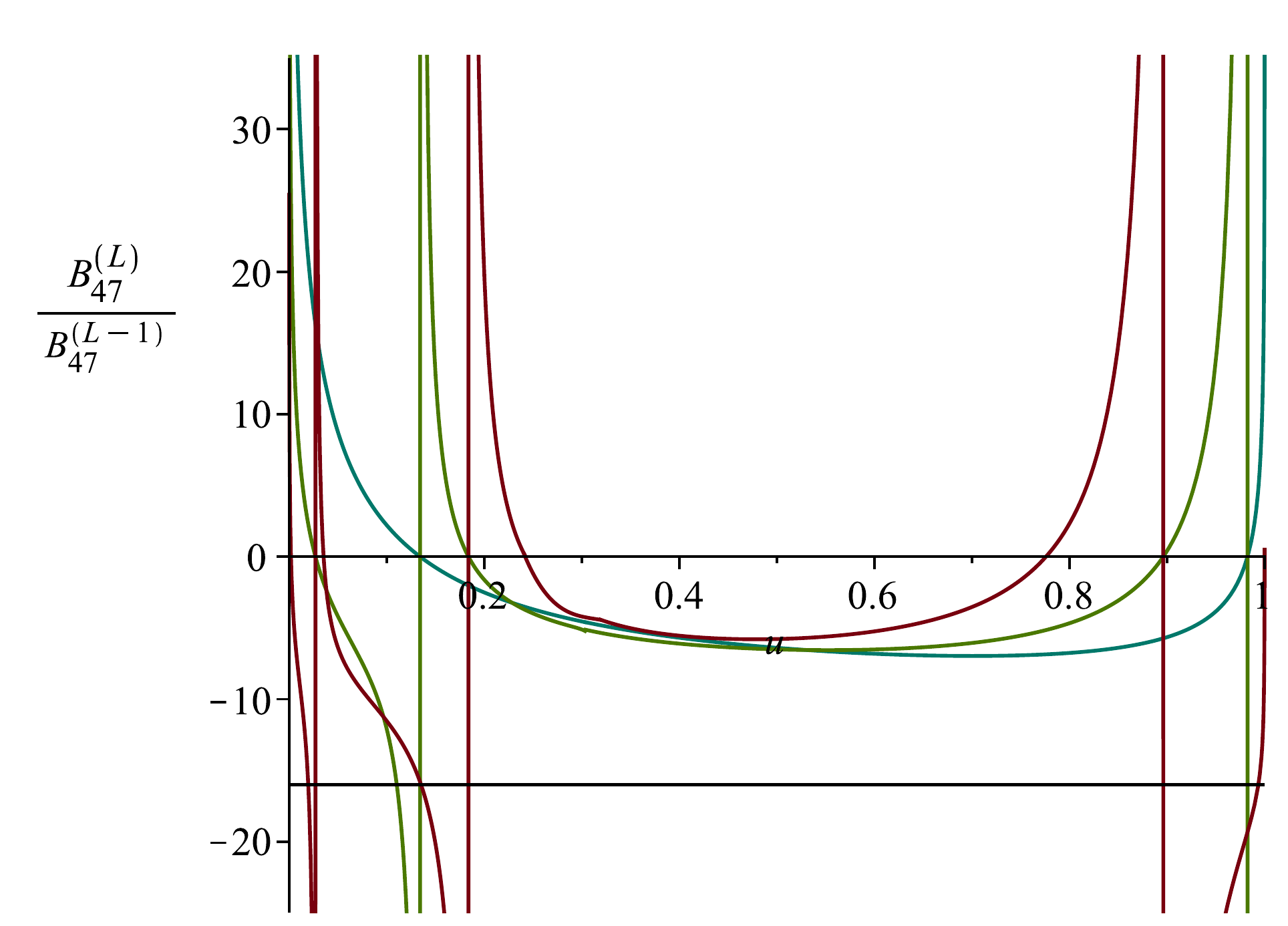}
\includegraphics[width=0.495\linewidth]{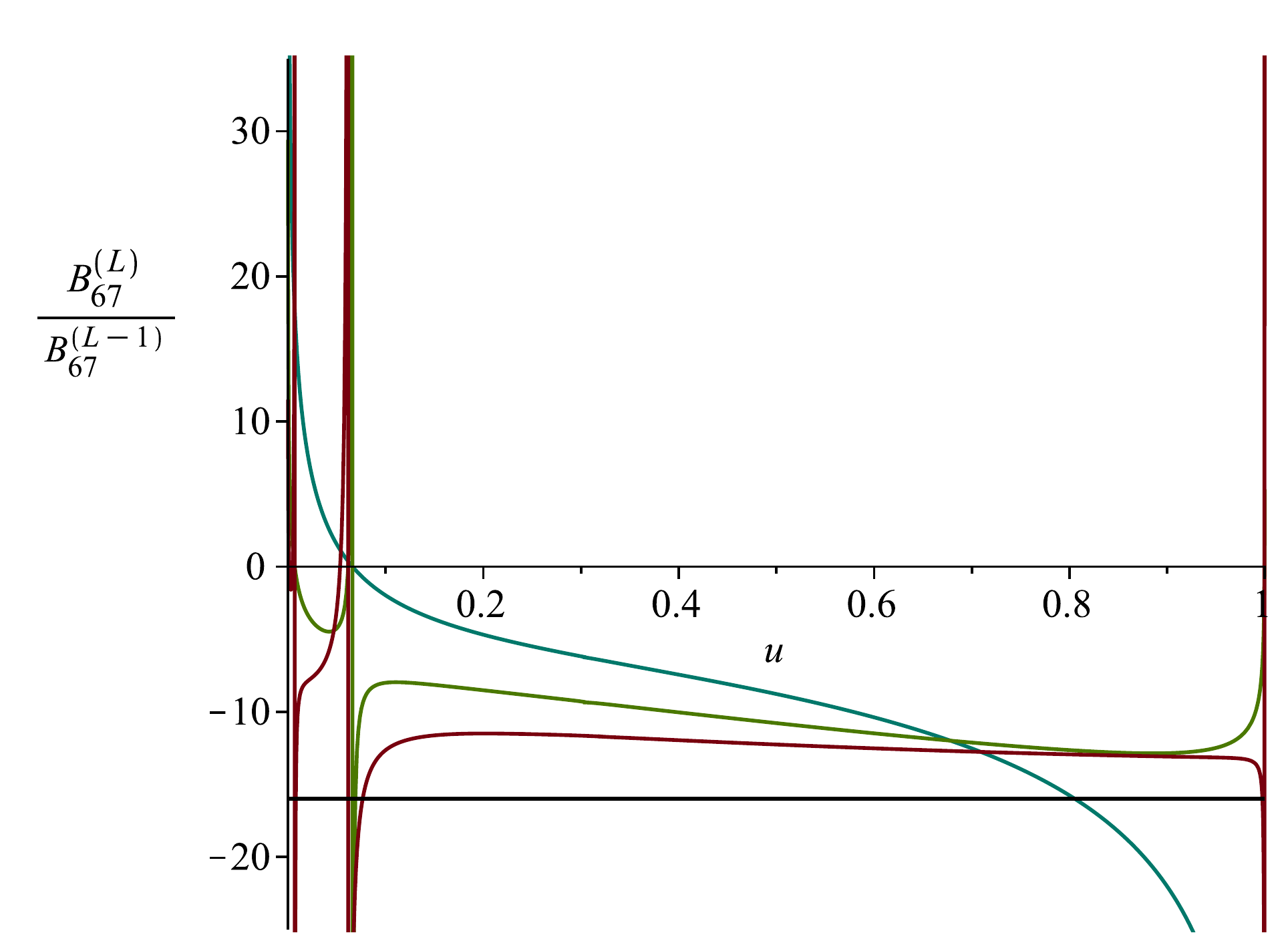}
\caption{Ratios of the BDS-normalized NMHV amplitude components $B_0$, $B_{47}$ and $B_{67}$ at successive loop orders on Line II (\ref{eq:line2}).}
\label{Fig:NMHVselfcross}
\end{figure}

In \Fig{Fig:NMHVselfcross}, we plot the ratios of BDS-normalized NMHV components at successive loop orders on Line II.  (These components have parity-odd pieces, and they would become imaginary on Line I where $\Delta\neq0$.)  The spikes in the plot occur when an $(L-1)$-loop amplitude crosses zero.  Usually the $L$-loop amplitude crosses zero nearby, so the spikes are near each other.  For $B_{47}$ they are a bit further apart.  For $u$ in the neighborhood of $0.5$, the ratios are relatively flat and are approaching the asymptotic cusp ratio value of $-16$, although $B_{47}$ is a bit less well-behaved in this respect too.

%%%%%%%%%%%%%%%%%%%%%%%%%%%%%%%%%%%%%%%%%%%%%%%
\section{The origin}
\label{sec:origin}

The heptagon origin is defined by $u_i \ll 1$, $i=1,2,\ldots,6$, $u_7=1$.
It sets the maximum number of two-particle invariants to zero, consistent
with the Gram determinant constraint~(\ref{eq:gram}).  It is a limiting case
of the CO surface.  In the six-point case, the logarithm of the MHV amplitude
at the hexagon origin is, remarkably, quadratic in the logarithms of the
cross ratios $u,v,w$ to all orders~\cite{Caron-Huot:2019vjl,Basso:2020xts}.
So it is of considerable interest to study the heptagon origin to see what happens there. Note that \eqn{eq:R72CO} for the two-loop remainder function $R_7^{(2)}$ is written in a form making it trivial to see its behavior at the origin, because only the last line survives (plus the exchange terms):
\bea
R_7^{(2)}(u_{1,\ldots,6}\ll 1) &=&
\z_2 \Bigl[ \ln u_1 \ln u_3 + \ln u_2 \ln u_4 + \ln u_3 \ln u_5 + \ln u_4 \ln u_6
\nonumber\\
&&\hskip0.4cm\null
+ \ln u_1 \ln u_2 + \ln u_2 \ln u_3 + \ln u_3 \ln u_4
    + \ln u_4 \ln u_5 + \ln u_5 \ln u_6 \Bigr]
\nonumber\\
&&\hskip0.1cm\null
+ \frac{17}{2} \, \z_4 \ + \ {\cal O}(u_i)\,.
\label{eq:R72origin}
\eea
If we take all the small cross ratios to be equal, $u_1 = u_2 = \cdots u_6 = u$,
\eqn{eq:R72origin} simplifies to
\be
R_7^{(2)}(u_{1,\ldots,6}=u\ll 1) =
9 \, \z_2 \, \ln^2 u + \frac{17}{2} \, \z_4 \ + \ {\cal O}(u)\,.
\label{eq:R72so}
\ee

Similarly, the three- and four-loop remainder functions are quadratic
polynomials in $\ln(u_i)$, with coefficients that are zeta values of
increasingly high weight.
Because the difference between $\ln \cE_7$ and $R_7$ is $\Gcusp \cE^{(1)}/4$,
which is manifestly quadratic in logarithms as well, the same will be
true of $\ln \cE_7$.  Indeed, this quadratic behavior can be
observed numerically in \Fig{Fig:lnEMHVsymline}: successive loop order ratios
for $\ln \cE_7$ on the symmetric diagonal line $(u,u,u,u,u,u,(1-u-u^2)^2/(1-2u^2))$ approach a constant as $u\to0$. A detailed investigation of the behavior
of the MHV amplitude at the origin will be
reported elsewhere~\cite{BDLPtoappear}.

In contrast to the MHV amplitude,
the six-point NMHV amplitude, or the ratio function,
was not observed to have any particularly simple behavior at the
hexagon origin, using data through six loops~\cite{Caron-Huot:2019vjl}.
The main property noted was that the ratio function components all
have maximum degree $L$ at $L$ loops, consistent with the leading OPE
behavior on the hexagon double-scaling surface.

Here we will report on the behavior of the seven-point NMHV amplitude
at the heptagon origin through four loops, although it also does not exhibit
any particular simple structure. We give more complete (and lengthy) formulas
for a generic approach to the origin in the ancillary file
\texttt{R\_P\_o\_co.txt}.
In the following, for compactness, we set $u_1 = u_2 = \cdots u_6 = u$.
Through four loops, the ``tree'' component $E_0$ behaves as,
\bea
E_0^{(1)} &=& - 6 \ln^2 u - 6 \z_2 \,,
\label{E01so}\\
E_0^{(2)} &=& \frac{99}{4} \ln^4 u + 82 \z_2 \ln^2 u - 18 \z_3 \ln u
+ \frac{367}{4} \z_4 \,,
\label{E02so}\\
E_0^{(3)} &=&
- \frac{517}{18} \ln^6 u - \frac{2483}{12} \z_2 \ln^4 u + 16 \z_3 \ln^3 u
- \frac{4899}{4} \z_4 \ln^2 u + (144 \z_5 + 138 \z_2 \z_3) \ln u
\nonumber\\&&\hskip0.1cm\null
- \frac{42671}{48} \z_6 - 33 (\z_3)^2 \,,
\label{E03so}\\
E_0^{(4)} &=& 
\frac{11987}{576} \ln^8 u + \frac{9275}{36} \z_2 \ln^6 u
+ \frac{421}{12} \z_3 \ln^5 u + \frac{158881}{48} \z_4 \ln^4 u 
- \Bigl( 181 \z_5 + \frac{599}{6} \z_2 \z_3 \Bigr) \ln^3 u
\nonumber\\&&\hskip0.1cm\null
+ \Bigl( \frac{1249343}{96} \z_6 + \frac{177}{2} (\z_3)^2 \Bigr) \ln^2 u
- \Bigl( 1350 \z_7 + 1272 \z_2 \z_5 + \frac{8261}{4} \z_4 \z_3 \Bigr) \ln u
\nonumber\\&&\hskip0.1cm\null
+ \frac{5266039}{576} \z_8 + 291 \z_2 (\z_3)^2 + 576 \z_3 \z_5 \,,
\label{E04so}
\eea
while the $E_{47}$ component is given by,
\bea
E_{47}^{(1)} &=& 2 \ln^2 u + \z_2 \,,
\label{E471so}\\
E_{47}^{(2)} &=& - \frac{9}{2} \ln^4 u - \frac{35}{2} \z_2 \ln^2 u
+ 6 \z_3 \ln u  - \frac{83}{4} \z_4 \,,
\label{E472so}\\
E_{47}^{(3)} &=& \frac{55}{18} \ln^6 u + \frac{383}{12} \z_2 \ln^4 u
- \frac{21}{2} \z_3 \ln^3 u + \frac{1093}{4} \z_4 \ln^2 u
- ( 48 \z_5 + 51 \z_2 \z_3 ) \ln u
\nonumber\\&&\hskip0.1cm\null
+ \frac{4073}{16} \z_6 + 10 (\z_3)^2 \,,
\label{E473so}\\
E_{47}^{(4)} &=& \frac{301}{96} \ln^8 u+ \frac{161}{8} \ln^6 u \z_2
+ \frac{7}{4} \z_3 \ln^5 u - \frac{19397}{96} \z_4 \ln^4 u
+ \Bigl( 102 \z_5 + \frac{292}{3} \z_2 \z_3 \Bigr) \ln^3 u
\nonumber\\&&\hskip0.1cm\null
- \Bigl( \frac{130369}{48} \z_6 + 24 (\z_3)^2 \Bigr) \ln^2 u
+ \Bigl( 450 \z_7 + 466 \z_2 \z_5 + \frac{3273}{4} \z_4 \z_3 \Bigr) \ln u
\nonumber\\&&\hskip0.1cm\null
- \frac{1742773}{576} \z_8 - 84 \z_2 (\z_3)^2 - 188 \z_3 \z_5 \,.
\label{E474so}
\eea

Notice the sign alternating behavior in $E_0^{(L)}$ from one loop order to the next.  In fact, taking into account that $\ln u$ is negative, the behavior is almost ``perfect'', term by term, the only exception being the $\z_3 \ln^5 u$ term
at four loops.  This behavior is reflected numerically in the $E_0$ ratio plot in \Fig{Fig:NMHVsymline} as $u\to0$.  There is a similar sign alternation in $E_{47}^{(L)}$ through three loops, but it fails utterly at four loops, which is also visible in in the $E_{47}$ ratio plot in \Fig{Fig:NMHVsymline}.

Also notice that at weight 8, the first irreducible MZV, $\z_{5,3}$, could potentially have appeared, but it does not appear, neither in eqs.~(\ref{E04so}) and (\ref{E474so}), nor in any of the other NMHV components or the MHV amplitude. For the hexagon origin, for both MHV and NMHV, this same absence of MZVs can be verified through seven loops, where the potential irreducible MZVs include $\z_{5,3}$, $\z_{7,3}$, $\z_{5,3,3}$, and several more at weights 12 to 14~\cite{Caron-Huot:2019vjl,DDtoappear}.

We have also inspected the 15 components of the ratio function,
\be \cP \equiv \cP^{(0)} \, \cP_0 + \big[ (12) \, \cP_{12} + (14) \, \cP_{14} + \text{ cyclic} \big]\,,
\label{eq:ratiofunctioncomponents}
\ee
at the heptagon origin.  We have verified that at $L$ loops (for $L\leq4$), all 15 components, $\cP_0$, $\cP_{i,i+1}$ and $\cP_{i,i+3}$, have maximum degree $L$ in all six logarithms $\ln u_i$ individually. The degree $L$ behavior is consistent with the general expected OPE behavior~\cite{Gaiotto:2011dt,Dixon:2011pw,Basso:2013vsa,Basso:2013aha,Caron-Huot:2019vjl}. On the full CO surface, the same maximum degree of $L$ also holds, with respect to the logarithms of the four small cross ratios, $\ln u_i$, $i=1,2,5,6$.

%%%%%%%%%%%%%%%%%%%%%%%%%%%%%%%%%
\section{Coaction and tropical fan comments}
\label{sec:coactioncomments}

\subsection{Amplitude coproducts and zeta values}
\label{sec:ampcop}

In the context of the hexagon function bootstrap it has proved very instructive to take an iterated $\{n-1,1\}$ coaction (essentially repeated derivatives) of the high loop order amplitudes, once they are determined. At each step of this ``top down'' analysis, one takes the linear span of the functions obtained, and then takes the $\{n-1,1\}$ coaction again.  The number of functions grows in the first few steps, but eventually it must shrink, in order to fit into the space constructed from the bottom up.  It may saturate the bottom-up space, in which case one has found the minimal function space needed to capture all the amplitudes (at least to the loop order computed).  Or there may be functions (perhaps starting with zeta-valued constants) that are not required by the amplitudes' derivatives, in which case one may consider removing them.  We have already begun this procedure in some sense, by not including $\z_2$ in the initial construction of the space, as mentioned in \sect{sec:bsnumber}.

There are a few options for how one carries out this procedure:  MHV and NMHV amplitudes could be analyzed separately or together. Also, parity even and odd functions could be treated together or separately.  Because there are 15 times as many NMHV amplitude components as MHV amplitudes at weight $2L$, there is not much difference between analyzing MHV and NMHV together, versus NMHV alone.  We will do MHV and NMHV together.  The generic NMHV component does not have definite parity.  However, if we consider the transcendental functions needed for both NMHV and the conjugate $\overline{\text{NMHV}}$ amplitudes together, then we should also consider the parity-even and parity-odd parts of the NMHV component functions to be separate functions.

In \Tab{tab:topdowneven} we show the number of independent weight $n$ parity-even $\{n,1,1,\ldots\}$ coproducts we get from applying this procedure to the $L$ loop MHV and NMHV amplitudes together. \Tab{tab:topdownodd} is the corresponding table for the parity-odd sector.

%%%%%%%%%%%%%%%%%%%%%%%%%%%%%%%5
\renewcommand{\arraystretch}{1.25}
\begin{table}[!t]
\centering
\begin{tabular}[t]{l c c c c c c c c c}
\hline\hline
weight $n$ & 0 & 1    &     2  &    3   &      4  &   5 &   6 & 7 &   8\\
\hline\hline
$L=1$ & \g{1} & \g{7} &    15  &        &         &     &     &   &    \\\hline
$L=2$ & \g{1} & \g{7} & \g{28} &    63  &     15  &     &     &   &    \\\hline
$L=3$ & \g{1} & \g{7} & \g{28} & \g{92} &    239  & 126 &  16 &   &    \\\hline
$L=4$ & \g{1} & \g{7} & \g{28} & \g{92} & \g{288} & 753 & 638 &
                                                       154 &  16 \\\hline\hline
\end{tabular}
\caption{Number of independent parity-even $\{n,1,1,\ldots,1\}$ coproducts of the MHV and NMHV seven-point amplitudes together through $L=4$ loops.  A green number denotes saturation of the space constructed from the bottom up.} \label{tab:topdowneven}
\end{table}
%%%%%%%%%%%%%%%%%%%%%%%%%%%%

%%%%%%%%%%%%%%%%%%%%%%%%%%%%%%%5
\renewcommand{\arraystretch}{1.25}
\begin{table}[!t]
\centering
\begin{tabular}[t]{l c c c c c c c c c}
\hline\hline
weight $n$ & 0 &   1  &  2    &    3  &    4   &     5   &  6   & 7 & 8\\
\hline\hline
$L=1$ & \g{0} & \g{0} & \g{0} &       &        &         &     &   &    \\\hline
$L=2$ & \g{0} & \g{0} & \g{0} & \g{6} &    14  &         &     &   &    \\\hline
$L=3$ & \g{0} & \g{0} & \g{0} & \g{6} & \g{28} &    118  &  15 &   &    \\\hline
$L=4$ & \g{0} & \g{0} & \g{0} & \g{6} & \g{28} & \g{120} & \red{406} &
                                                        154 &  15 \\\hline\hline
\end{tabular}
\caption{Number of independent parity-odd $\{n,1,1,\ldots,1\}$ coproducts of the MHV and NMHV seven-point amplitudes together through $L=4$ loops.  A green number denotes saturation of the space constructed from the bottom up.
The $\red{406}$ is discussed in the text.} \label{tab:topdownodd}
\end{table}
%%%%%%%%%%%%%%%%%%%%%%%%%%%%

By the time we include the four-loop amplitudes, the entire space is saturated through weight 4.  This includes the independent constants $\z_3$ and $\z_4$.  As remarked in \sect{sec:bsnumber}, it is enough to know that the 120 weight 5 parity-odd functions are present (as indicated by \Tab{tab:topdownodd}) to conclude that $\z_3$ must be an independent function.

From \Tab{tab:bottomup}, the first time we can have parity-odd beyond-the-symbol functions is at weight 6, corresponding to multiplying the 6 weight 3 parity-odd functions (the one-mass scalar hexagon integrals in six dimensions) by $\z_3$. That would add to the 406 weight 6 parity-odd symbols to give an expected 412 functions.  However, we only find 406 independent odd functions among the $\{6,1,1\}$ coproducts of the four-loop amplitude.  This might be an accident, i.e.~it might be completed to all 412 when the same analysis is done at five loops.  On the other hand, the 406 independent functions are not arbitrary; they each correspond to one of the 406 symbol-level functions, plus a specific linear combination of the 6 beyond-the-symbol functions. (We provide these combinations in the ancillary file \texttt{weight6odd406.txt}.) In other words, it may be that $\z_3$ should not be considered independent with respect to the parity-odd sector.  Five loop data would be welcome to address this question.

We computed the $\{4,1,1\}$ double coproducts of the 406 weight 6 odd functions and found that the entire weight 4 space was in its span, including $\z_4$.  Recall that the analogous statement was also true at one lower weight.  It would be interesting to construct just the parity-odd sector at weight 7, in order to see if $\z_5$ and $\z_2 \z_3$ are contained in its $\{5,1,1\}$ double coproduct span.  At the moment we don't know whether these two constants should be considered independent or not.  They are not contained in the 753-dimensional span of parity-even $\{5,1,1,1\}$ coproducts of the four-loop amplitude.  Our suspicion, from the example of $\z_3$, is that they should be independent constants, at least for the parity-even sector.

From the point of view of continuity under $7\to6$ soft or collinear limits, it is somewhat puzzling that $\z_3$ needs to be independent in the heptagon function space, while it was not required to be for the hexagon functions.  (Both hexagon and heptagon function spaces allow $\z_2$ to be fixed, and require $\z_4$ to be independent.)  Of course many heptagon functions (both even and odd) blow up logarithmically in soft and collinear limits, and this discontinuous behavior may be involved in resolving this issue.

In any event, the existence of $\z_3$ with a free parameter seems to imply that the coaction principle is less powerful for heptagon functions than for hexagon functions.  For example, the functions $\z_3 \, \ln u_i$ also must appear with independent coefficients and presumably a large tower of higher-weight functions. In ref.~\cite{Caron-Huot:2019bsq} some of the most striking consequences of the coaction principle were associated with the values of functions at points, especially the dihedrally symmetric bulk base point $(u,v,w)=(1,1,1)$ where all hexagon functions evaluated to MZVs.  We don't know of an analog of this point for heptagon functions, so it is hard to search for similar restrictions at specific points.

The closest analog to the hexagon $(1,1,1)$ point might be the heptagon origin, even though it is far from the bulk.  When we take the odd-weight parity-odd functions to the origin, we do find an absence of odd zeta values in the constant terms. This statement is not true for the parity-even functions.  Specifically, there is no $\z_3$ in the limiting behavior of any of the 6 weight 3 parity-odd functions, and there is no $\z_5$ in the limiting behavior of the 120 weight 5 parity-odd functions, although $\z_2\z_3$ can be present.  The constant $(\z_3)^2$ is present in the limits of the weight 6 parity-odd functions, and this statement is independent of the 6 beyond-the-symbol functions, because they have vanishing $(\z_3)^2$ coefficients.  At weight 7, we don't have a full parity-odd basis, but there are 154 independent odd functions among the four-loop $\{7,1\}$ coproducts, as shown in \Tab{tab:topdownodd}, and there is no $\z_7$ in their limiting behavior at the origin, although both $\z_2\z_5$ and $\z_4\z_3$ are present there. In summary, there is an indication of constrained zeta values at the heptagon origin in the parity-odd sector, which is not present in the even sector.

We recall that in the hexagon function space, starting at weight 8, a handful of \emph{dropout functions} do not appear, even though their symbols pass all of the symbol-level requirements.  (See Table 10 of ref.~\cite{Caron-Huot:2019bsq}.) So far we have no evidence for this phenomenon in the heptagon function space.  However, it seems to be linked to the removal of independent constant zeta values at lower weights, which is more prevalent in the hexagon case; and besides, we currently have very little information about the heptagon function space above weight 6. 

%%%%%%%%%%%%%%%%%%%%%%%%%
\subsection{Tropical fans at function level}
\label{sec:tropicalfans}

In ref.~\cite{Drummond:2017ssj} it was observed that the symbols of the MHV amplitudes through four loops satisfied a pair adjacency condition that went beyond cluster adjacency.  The condition, promoted to function level, can be phrased as
\be\label{eq:a3fanconstraint}
F^{a_{21},a_{64}}\ =\ 0, \qquad \text{plus dihedral images.}
\ee
Here $F$ can be an MHV amplitude or any of its $\{n,1,\ldots,1\}$ coproducts.  (It could even correspond to arbitrary adjacent slots of the full coaction, but we don't have as easy access to this information beyond the $\{n,1,\ldots,1\}$ case.) As indicated, there are really 14 such conditions, obtained by applying the dihedral group $D_7$. Later these conditions were associated with edges in a certain ``$\{a_1,a_2,a_3\}$ tropical fan'' for the Grassmannian Gr(3,7)~\cite{Drummond:2019cxm,Arkani-Hamed:2019rds,Henke:2019hve,Drummond:2020kqg}.

We can now ask whether the conditions~(\ref{eq:a3fanconstraint}) are obeyed at full function level.  The answer is that they are, for MHV amplitudes through four loops. In fact, the conditions are obeyed for \emph{every} function in the heptagon function space through weight 5, and at weight 6 for all but a \emph{unique} parity-even function.\footnote{This fact also means that the conditions are obeyed by the pure beyond-the-symbol functions through weight 8, because they are constructed by multiplying $\z_3$ by functions of at most weight 5.}  Thus, the first place they could be violated (working from the left to the right in the symbol entries) is beginning in the 5-6 pair of slots, for parity-even $\{6,1,\ldots,1\}$ coproducts of amplitudes.

Actually, the conditions~(\ref{eq:a3fanconstraint}) are also obeyed by the NMHV amplitude through three loops, but are violated by the NMHV four-loop amplitude~\cite{Henke:2019hve,Drummond:2020kqg}.  The violation starts in the 5-6 slot, and in that slot it has a simple form, in that the violations in all components are related to each other,
\bea\label{eq:Eijnonzero}
E_{51}^{a_{21},a_{64},x,y} &=& E_{73}^{a_{21},a_{64},x,y} = E_{71}^{a_{21},a_{64},x,y} 
= E_{23}^{a_{21},a_{64},x,y} = E_{56}^{a_{21},a_{64},x,y} = - E_{0}^{a_{21},a_{64},x,y} \,,
\\
E_{i,i+1}^{a_{21},a_{64},x,y} &=& E_{i,i+3}^{a_{21},a_{64},x,y} = 0,
\quad \text{otherwise,}
\label{eq:Eijzero}
\eea
where $x$ and $y$ are arbitrary letters.  The violation in the 6-7 slot is more intricate, involving seven independent weight 5 functions.  There is no violation of \eqn{eq:a3fanconstraint} in the 7-8 slot for four-loop NMHV.

There is also a parity conjugate set of conditions to
\eqn{eq:a3fanconstraint}, namely
\be\label{eq:a3fanconstraintconjugate}
F^{a_{31},a_{65}}\ =\ 0, \qquad \text{plus dihedral images.}
\ee
Everything that is true for the first set is also true for this set,
except that the violation of it by the four-loop NMHV amplitude looks
slightly different (because the NMHV components are not invariant
under parity, instead they transform into $\overline{\text{NMHV}}$
components).  For the violation in the 5-6 slot, there is still a
unique weight 4 function involved,
but instead of eqs.~(\ref{eq:Eijnonzero}) and (\ref{eq:Eijzero})
we find the same relations cycled forward by one unit,
\bea\label{eq:Eijnonzeroconj}
E_{62}^{a_{31},a_{65},x,y} &=& E_{14}^{a_{31},a_{65},x,y} = E_{12}^{a_{31},a_{65},x,y} 
= E_{34}^{a_{31},a_{65},x,y} = E_{67}^{a_{31},a_{65},x,y} = - E_{0}^{a_{31},a_{65},x,y} \,,
\\
E_{i,i+1}^{a_{31},a_{65},x,y} &=& E_{i,i+3}^{a_{31},a_{65},x,y} = 0,
\quad \text{otherwise.}
\label{eq:Eijzeroconj}
\eea
The violation in the 6-7 slot for $a_{31},a_{65}$
involves only one independent function, not the seven found
for $a_{21},a_{64}$.  Again there is no violation in the 7-8 slot.

Finally, there is a third adjacency condition,
\be\label{eq:thirdconstraint}
F^{a_{11},a_{41}}\ =\ 0, \qquad \text{plus dihedral images,}
\ee
which has been observed to be obeyed by the MHV amplitudes' symbols
through four loops, but violated by the four-loop NMHV amplitude.
It goes beyond the constraints from the $\{a_1,a_2,a_3\}$ fan.
We have verified that this condition also holds at full function level
for the MHV case.  Again \emph{every} function in the heptagon function
space through weight 5 obeys \eqn{eq:thirdconstraint},
and at weight 6 all but a \emph{unique} function obeys it; however,
in this case the function does not have definite parity.
Again the violation for the four-loop NMHV amplitude starts in the 5-6 slot,
and takes almost as simple a form as eqs.~(\ref{eq:Eijnonzero}) and
(\ref{eq:Eijzero}):
\bea
E_{71}^{a_{11},a_{41},x,y} &=& E_{51}^{a_{11},a_{41},x,y} \,, \qquad\quad
E_{12}^{a_{11},a_{41},x,y} = E_{14}^{a_{11},a_{41},x,y} \,,
\label{eq:EijthirdnonzeroA}\\
E_{36}^{a_{11},a_{41},x,y} &=& E_{34}^{a_{11},a_{41},x,y} = E_{56}^{a_{11},a_{41},x,y}
= - E_{0}^{a_{11},a_{41},x,y}
= E_{51}^{a_{11},a_{41},x,y} + E_{14}^{a_{11},a_{41},x,y} \,,~~~
\label{eq:EijthirdnonzeroB}\\
E_{i,i+1}^{a_{11},a_{41},x,y} &=& E_{i,i+3}^{a_{11},a_{41},x,y} = 0,
\quad \text{otherwise.}
\label{eq:Eijthirdzero}
\eea
The violation of \eqn{eq:thirdconstraint} in the 6-7 slot involves
two independent functions, while again there is no violation in the
7-8 slot.

While the four-loop NMHV amplitude exhausts all the adjacent pairs
of letters allowed by cluster adjacency, at the level of triplets
its symbol falls seven short of the number allowed, corresponding to the
absence of 
\be\label{eq:a5fanconstraint}
F^{a_{11},a_{41},a_{51}}\ =\ 0, \qquad \text{plus dihedral images,}
\ee
and this absence is associated with triangles (rather than edges)
in a larger $\{a_1,\ldots,a_5\}$ fan~\cite{Drummond:2020kqg}.
Again we have verified that \eqn{eq:a5fanconstraint} holds at full
function level for the NMHV amplitude through four loops
(and of course also for MHV).  We also checked it for the parity conjugate
triplet, $a_{11},a_{51},a_{41}$.

In summary, the MHV edge constraints associated with the $\{a_1,a_2,a_3\}$ tropical fan, the additional MHV constraints~(\ref{eq:thirdconstraint}), and the NMHV triangular constraints associated with the $\{a_1,\ldots,a_5\}$ tropical fan, first seen at symbol level, all hold as well at full function level. This fact may bode well for the significance of tropical fans for higher-point amplitudes. On the other hand, since the ability to violate these constraints develops fairly late in the available slots, it is fair to wonder whether they are an accident of only having data through four loops.  It would certainly be of interest to test the tropical fan predictions for seven-point amplitudes at five loops.

%%%%%%%%%%%%%%%%%%%%%%%%%%%%%%%%%%%%%%%%%%%%%%%
\section{Conclusions and outlook}
\label{sec:conclusions}

In this paper we have lifted the Steinmann heptagon cluster boostrap program~\cite{Drummond:2014ffa,Dixon:2016nkn,Drummond:2018caf} from the level of symbols to the level of functions.  In particular, we have computed the MHV and NMHV seven-point amplitudes through four loops, the same order at which their symbols were known previously.  We have also achieved a complete function basis through weight six.

We did so by making use of a triple-scaling ``CO'' surface that interpolates between soft and collinear limits and the heptagon origin, and on which the heptagon functions drastically simplify.  This surface makes it simple to implement branch cut constraints, as well as the soft and collinear limits that can be used to constrain amplitudes at the full function level.  We expect that similar surfaces will prove quite useful beyond seven points~\cite{BDLPtoappear}.

The (Euclidean) CO surface lets us fix boundary conditions for evaluating the amplitudes in any other region that we can reach from it.  As an example, we plotted the amplitudes on a few lines entering the Euclidean bulk, which touch the CO surface at one end.  In some cases, such as the Euclidean self-crossing Line II, the integration can be done analytically; in other cases we integrated up using a high-order series expansion.

There are many other kinematic regions we can investigate in the future.  For example, the OPE limit is predicted to arbitrary loop order and to high orders in the OPE in the Pentagon Operator Product Expansion program~\cite{Basso:2013vsa,Basso:2013aha,Basso:2014koa}.  Many tests have been done already for the six-point amplitudes (see e.g.~\cite{Caron-Huot:2019vjl}), but there the pentagon transition probed has excitations only on one side of the pentagon.  In seven-point amplitudes one can test the full pentagon transitions, with excitations on both sides.  The OPE limit sends $u_2 = T_2^2$ and $u_5 = T_1^2$ to zero. At leading order in the limit $u_3=(1-u_1)/(1-u_1u_6)$, $u_4=(1-u_6)/(1-u_1u_6)$, $u_7=1-u_1u_6$.  It intersects the CO surface when $u_1,u_6 \to 0$, i.e.~at $(u_3,u_4)=(1,1)$. We have checked the NMHV scalar pentagon transition at this leading level through four loops~\cite{Basso:2013aha,Bassoprivate}, but many more detailed checks are possible.

As another example, when a three-particle Mandelstam invariant vanishes, the seven-point amplitude has a multi-particle factorization into the product of a four-point and a five-point amplitude.  This limit was studied at symbol level~\cite{Dixon:2016nkn} but the complete function-level behavior should also be explored.

Other limits require leaving the Euclidean sheet to arrive at physical $2\to5$ or $3\to4$ scattering.  The self-crossing limit on such a physical sheet is one important limit, which mimics double parton scattering~\cite{Dixon:2016epj}.

Another physical scattering limit which has received much attention is the MRK limit.  Recently, an all-order proposal has been made~\cite{DelDuca:2019tur} for the central emission block, which first appears at the seven-point level.  It will be important to test this proposal by analytically continuing the full function level amplitude into this region.  (Certain checks have already been done~\cite{DelDuca:2018hrv} at next-to-leading logarithmic accuracy, based on completing the symbol to a function just in the MRK limit.)  The MRK limit can be taken by letting $u_{3,4,7}\to1$, $u_{1,2,5,6}\to0$, holding fixed $u_1/(1-u_3)$, $u_2/(1-u_3)$, $u_5/(1-u_4)$, and $u_5/(1-u_4)$~\cite{Bartels:2011ge,Broedel:2016kls}.  Thus the MRK limit touches the CO surface at $(u_3,u_4)=(1,1)$.  To be sensitive to the ``long'' Regge cut, which is sensitive to the central emission block, one must first analytically continue $u_7 \to u_7 e^{-2\pi i}$.  This analytic continuation can also be performed with the aid of the CO surface, since $u_3$ and $u_4$ are generic and can be moved near the origin. However, since $u_7=1$ on our reference CO surface, we must use the cyclic symmetry~(\ref{eq:COcycle}) to exploit a permuted CO surface in which $u_7$ is generic instead.  After analytically continuing, we can transport integration constants for the correct sheet back to the desired CO surface. A similar analytic continuation will be needed to study the physical self-crossing limit.

We also studied the coaction principle for heptagon functions.  In contrast to the hexagon function case, $\z_3$ is an independent constant function, which limits the power of the coaction principle somewhat.  Nevertheless, we found some indication of zeta-value constraints in the parity-odd sector.

In conclusion, the heptagon function bootstrap is now well underway, and some of the methods employed seem likely to be very useful for many other circumstances, including yet higher-point amplitudes in planar $\cN=4$ super-Yang-Mills theory.

\acknowledgments
We would like to thank Benjamin Basso, Andrew McLeod and Georgios Papathanasiou for collaboration on related projects and for very illuminating discussions, Enrico Herrmann for stimulating remarks, and Andrew McLeod, Mark Spradlin and Georgios Papathanasiou for useful comments on the manuscript. We are also grateful to Benjamin Basso for OPE assistance. This research was supported by the US Department of Energy under contract DE--AC02--76SF00515. YL acknowledges support from the Benchmark Stanford Graduate Fellowship. LD thanks the Donner Lake Institute for Theoretical Physics for hospitality while this research was completed.

%%%%%%%%%%%%%%%%%%%%%%%%%%%%%%%%%%%%%%%%%%%%%%%%%%%%%%%%%%%

\appendix
\section{BDS ansatz}  \label{app:bds}

Before turning to the BDS ansatz, we record for convenience the cross ratios $u_i$, defined in \eqn{eq:uidef}, in terms of Mandelstam invariants $s_{i,i+1} = (p_i+p_{i+1})^2$ and $s_{i,i+1,i+2} = (p_i+p_{i+1}+p_{i+2})^2$:
\bea
u_1 &=& \frac{s_{34}s_{671}}{s_{234}s_{345}} \,, \qquad
u_2 = \frac{s_{45}s_{712}}{s_{345}s_{456}} \,, \qquad
u_3 = \frac{s_{56}s_{123}}{s_{456}s_{567}} \,, \qquad
u_4 = \frac{s_{67}s_{234}}{s_{567}s_{671}} \,, \nonumber\\
u_5 &=& \frac{s_{71}s_{345}}{s_{671}s_{712}} \,, \qquad
u_6 = \frac{s_{12}s_{456}}{s_{712}s_{123}} \,, \qquad
u_7 = \frac{s_{23}s_{567}}{s_{123}s_{234}} \,. 
\label{eq:uisi}
\eea

The BDS ansatz \cite{Bern:2005iz} captures the infrared divergences of the amplitudes. In dimensional regularization, the $n$ particle BDS ansatz is
\be
\cA_n^{\text{BDS}} = \cA_n^{\text{MHV,tree}}\cdot \exp\left[\sum_{L=1}^\infty g^{2L}\left(f^{(L)}(\eps)\cdot M_n(L\eps)\, +\, \text{const.} \right) \right],
\ee
where
\be
f(\eps) = \sum_{L=1}^\infty g^{2L} f^{(L)}(\eps) = \frac{\Gcusp}{4} \, + \, O(\eps),
\ee
and $M_n$ is the one-loop amplitude normalized by the tree. In the case of seven particles~\cite{Bern:1994zx,Dixon:2016nkn},
\be
M_7(\eps) = - \frac{1}{\eps^{2}} \sum_{i=1}^{7} \left( \frac{\mu^{2}}{-s_{i,i+1}} \right)^{\eps} + F_7+\mathcal{O}(\eps),
\ee
where
\bea
F_7 &=& \sum_{i=1}^{7} \biggl[ \Li_{2} \left(1{-}\frac{1}{u_{i}} \right)
  + \frac{1}{2} \ln\left(\frac{u_{i+2}u_{i{-}2}}{u_{i+3}u_{i}u_{i{-}3}}\right)
  \ln u_{i} \nonumber\\
&&\hskip1cm\null
  + \ln(s_{i,i+1}) \ln\left( \frac{s_{i,i+1}s_{i+3,i+4}}{s_{i+1,i+2}s_{i+2,i+3}} \right)
  + \frac{3}{2} \zeta_{2} \biggr].
\eea
We see from the above formula that the additional factor
\be
\exp\left[-\frac{\Gcusp}{4} \, \cE_7^{(1)} \right]
\ee
in the BDS-like ansatz (\ref{eq:bdslike}) removes its dependence on the three-particle Mandelstam variables, so that BDS-like normalized amplitudes will obey the three-particle Steinmann relations.

%%%%%%%%%%%%%%%%%%%%%%%%%%%%%%%%%

\section{Symbol alphabet on the Collinear-Origin surface}
\label{app:symco}

Using the expressions~(\ref{gletters}) for the 42 $g$-letters in terms of the $a$-letters, the representation~(\ref{aletters}) for the $a$-letters in terms of momentum-twistor four-brackets, and the parametrization~(\ref{eq:TSF7}) of the momentum twistors, one can take the triple scaling limit~(\ref{eq:tslimit}) of the $g$-letters.  In particular, the limits of the cross ratios $g_{1i} \equiv u_i$ are given by \eqn{eq:tsui}.  One can rewrite the limits of the remaining 35 letters in terms of $u_1$ to $u_6$, and the symbol alphabet on the Collinear-Origin surface collapses to just 9 letters,
\be
u_1, \quad u_2, \quad u_3, \quad 1-u_3, \quad u_4, \quad 1-u_4, \quad u_5, \quad u_6, \quad 1-u_7,
\ee
as follows,
\vspace{.1cm}
\begin{footnotesize}
\begin{align}
g_{11} &= u_1,& g_{21} &= 1,&
g_{31} &= 1,& g_{41} &= 1-u_4,&
g_{51} &= \frac{(1-u_4)^2}{u_3 u_4 u_5 u_6},&
g_{61} &= \frac{u_6 (1-u_3)^2}{u_1 u_2 u_3 (1-u_4)},&
\nonumber\\[1ex]
g_{12} &= u_2,& g_{22} &= 1,&
g_{32} &= 1-u_4,& g_{42} &= 1,&
g_{52} &= \frac{u_4 u_5 u_6}{(1-u_4)^2},&
g_{62} &= \frac{u_1 u_2 u_3 u_4}{(1-u_3)^2},&
\nonumber\\[1ex]
g_{13} &= u_3,& g_{23} &= 1-u_3,&
g_{33} &= 1,& g_{43} &=1-u_4,&
g_{53} &= \frac{u_1 (1-u_4)^2}{u_5 u_6 (1-u_3)},&
g_{63} &= \frac{(1-u_3)^2(1-u_4)}{u_1 u_2 u_3 u_4 u_5},&
\nonumber\\[1ex]
g_{14} &= u_4,& g_{24} &= 1-u_4,&
g_{34} &= 1,& g_{44} &= 1-u_3,&
g_{54} &= \frac{u_6 (1-u_3)^2}{u_1 u_2 (1-u_4)},&
g_{64} &= \frac{u_2 u_3 u_4 u_5 u_6}{(1-u_3)(1-u_4)^2},&
\nonumber\\[1ex]
g_{15} &= u_5,& g_{25} &= 1,&
g_{35} &= 1-u_3,& g_{45} &= 1,&
g_{55} &= \frac{u_1 u_2 u_3}{(1-u_3)^2},&
g_{65} &= \frac{(1-u_4)^2}{u_3 u_4 u_5 u_6},&
\nonumber\\[1ex]
g_{16} &= u_6,& g_{26} &= 1,&
g_{36} &= 1,& g_{46} &= 1-u_3,&
g_{56} &= \frac{(1-u_3)^2}{u_1 u_2 u_3 u_4},&
g_{66} &= \frac{u_4 u_5 u_6 (1-u_3)}{u_1 (1-u_4)^2},&
\nonumber\\[1ex]
g_{17} &= 1,& g_{27} &= 1-u_7,&
g_{37} &= 1,& g_{47} &= 1,&
g_{57} &= \frac{u_2 u_3 u_4 u_5 (1-u_7)}{(1-u_3)^2(1-u_4)^2},&
g_{67} &= \frac{u_1 u_2(1-u_4)^2}{u_5 u_6 (1-u_3)^2}.
\nonumber\\[1ex]
~&~ \label{eq:COgletters}
\end{align}
\end{footnotesize}
This table determines the derivative of any function $F$ with respect to every cross ratio along the CO surface.  For example,
\be\label{eq:COu1deriv}
u_1 \frac{\partial F}{\partial u_1} = F^{g_{11}}
+ F^{g_{53}} - F^{g_{54}} + F^{g_{55}} - F^{g_{56}}
- F^{g_{61}} + F^{g_{62}} - F^{g_{63}} - F^{g_{66}} + F^{g_{67}} \,.
\ee
One might expect another contribution to \eqn{eq:COu1deriv} from the
dependence on $u_1$ inside $(1-u_7)$, according to \eqn{eq:COGramdet}.
However, the $(1-u_7)$ branch cut condition implies that on the CO surface,
\be\label{eq:constraint2757}
F^{g_{27}} + F^{g_{57}} = 0.
\ee
Hence one can ignore the two $(1-u_7)$ factors in the table when computing derivatives.

%%%%%%%%%%%%%%%%%%%%%%%%

\section{Simple soft and collinear limits}
\label{app:simplesoftcollinear}

A particularly simple version of the soft limit $p_1\to0$ can be obtained from the parametrization~(\ref{eq:TSF7}) by letting $T_2 = -S_2$, $F_2=1$, $F_1 = -T_1S_1$, $T_1 \to \eps \cdot T_1$, $S_1 \to S_1/\eps$, and sending $\eps \to 0$. We further assume that $T_1S_1 \ll 1$.  Then there are two infinitesimal cross ratios, $u_5$ and $u_6$, and one generic one, $u_2$.  Following the same steps as for the CO surface, we obtain for the $g$-letters in this limit,
\vspace{.1cm}
\begin{footnotesize}
\begin{align}
g_{11} &= 1,& g_{21} &= \text{const.},&
g_{31} &= 1,& g_{41} &= u_2 u_6,&
g_{51} &= \text{const.},&
g_{61} &= u_2,&
\nonumber\\[1ex]
g_{12} &= u_2,& g_{22} &= 1-u_2,&
g_{32} &= u_2 u_6,& g_{42} &= 1,&
g_{52} &= \frac{u_6(1-u_2)}{u_5},&
g_{62} &= 1,&
\nonumber\\[1ex]
g_{13} &= u_2,& g_{23} &= 1-u_2,&
g_{33} &= 1,& g_{43} &= u_5 u_6 (1-u_2),&
g_{53} &= \frac{u_5}{u_6 (1-u_2)},&
g_{63} &= \frac{u_5^2}{u_6 (1-u_2)},&
\nonumber\\[1ex]
g_{14} &= 1,& g_{24} &= u_2 u_6,&
g_{34} &= 1,& g_{44} &= 1-u_2,&
g_{54} &= 1,&
g_{64} &= \frac{u_6 (1-u_2)}{u_5},&
\nonumber\\[1ex]
g_{15} &= u_5,& g_{25} &= 1,&
g_{35} &= 1-u_2,& g_{45} &= u_5 u_6^2 u_2 (1-u_2),&
g_{55} &= 1,&
g_{65} &= \frac{u_5^2}{u_6 (1-u_2)},&
\nonumber\\[1ex]
g_{16} &= u_6,& g_{26} &= 1,&
g_{36} &= u_2 u_6,& g_{46} &= 1-u_2,&
g_{56} &= 1,&
g_{66} &= \frac{u_6 (1-u_2)}{u_5},&
\nonumber\\[1ex]
g_{17} &= 1,& g_{27} &= u_5 u_6 (1-u_2),&
g_{37} &= 1,& g_{47} &= u_5 u_6 u_2,&
g_{57} &= \frac{u_6 (1-u_2)}{u_5^2},&
g_{67} &= 1,
\nonumber\\[1ex]
~&~ \label{eq:soft1vvgletters}
\end{align}
\end{footnotesize}
where ``const.'' means a letter is singular but it does not contribute to any derivative along the surface.  The six point cross ratios are $(u,v,w)=(1,u_2,u_2)$, which is a line along which the hexagon functions are HPLs $H_{\vec{w}}(u_2)$ with $w_k \in \{0,1\}$.  Here such functions are tensored with logarithms in the small variables $u_5$ and $u_6$. All parity-odd functions vanish in this limit. We can impose the branch-cut condition
\be\label{eq:softbranchcut}
F^{1-u_2} = 0 \quad\quad \text{when } u_2 \to 1\,,
\ee
on this one-dimensional line.  This condition provides additional restrictions beyond the CO surface conditions~(\ref{eq:CObranchcut}).  Because $u_1=1$, this soft limit does not touch the CO surface directly, but it can be reached via an intermediate collinear limit, which we now describe.

A particularly simple version of a collinear limit follows from the parametrization~(\ref{eq:TSF7}) by first letting $T_j \to T_j \cdot \eps$.  This is the standard OPE limit, with infinitesimal letters $u_2 = T_2^2$ and $u_5 = T_1^2$, and generic letters $u_1$, $1-u_1$, $u_6$, $1-u_6$ and $1-u_1u_6$. To simplify things further, we will take $u_6\to0$.  The $g$-letters then become,
\vspace{.1cm}
\begin{align}
g_{11} &= u_1,& g_{21} &= 1-u_1,&
g_{31} &= 1,& g_{41} &= u_6,&
\nonumber\\[1ex]
g_{12} &= u_2,& g_{22} &= 1,&
g_{32} &= u_6,& g_{42} &= 1,&
\nonumber\\[1ex]
g_{13} &= 1-u_1,& g_{23} &= u_1,&
g_{33} &= 1,& g_{43} &= u_6,&
\nonumber\\[1ex]
g_{14} &= 1,& g_{24} &= u_6 (1-u_1),&
g_{34} &= 1,& g_{44} &= u_1,&
\nonumber\\[1ex]
g_{15} &= u_5,& g_{25} &= 1,&
g_{35} &= u_1,& g_{45} &= 1-u_1,&
\nonumber\\[1ex]
g_{16} &= u_6,& g_{26} &= 1,&
g_{36} &= 1-u_1,& g_{46} &= u_1,&
\nonumber\\[1ex]
g_{17} &= 1,& g_{27} &= u_6 u_1,&
g_{37} &= 1,& g_{47} &= 1-u_1.&
\label{eq:collgletters}
\end{align}
All the odd letters $g_{5,j}$, $g_{6,j}$ depend on the $F_j$ but they can be neglected at leading order in this limit, because the parity-odd functions vanish.  Functions on this collinear line collapse to HPLs $H_{\vec{w}}(u_1)$ with $w_k \in \{0,1\}$, multiplied by logarithms in $u_2$, $u_5$, and $u_6$.

Because $u_3=1-u_1$, there is no branch cut condition on this line.  But it interpolates between the above soft limit and the CO surfaces.  The limit of the collinear line where $u_3=1-u_1\to0$, after letting $u_3 = u_2$, matches the $u_2\to0$ limit of the above soft limit.  Whereas the limit $u_1\to0$ of the collinear line matches the $(u_3,u_4) \to (1,1)$ limit of the CO surface. Thus one can use the collinear line to transport the soft branch cut condition~(\ref{eq:softbranchcut}) back to the CO surface.

%%%%%%%%%%%%%%%%%%%%%%%%

\section{General soft limit in terms of momentum twistors}
\label{app:softcollinear}

Here we discuss how the general soft limit $p_1\to0$ onto arbitrary hexagon kinematics can be taken using momentum twistors~\cite{DelDuca:2016lad}. Since $Z_i\in\mathbb{P}^3$, in homogeneous coordinates, $Z_1$ can be written as a linear combination\footnote{One might be concerned that the coefficients $c_i$ in \eqn{eq:Z1combo} carry little group weights; this does not cause problems since the little group weights always cancel in Lorentz invariant quantities.
}
\be\label{eq:Z1combo}
Z_1 = c_6 Z_6 + c_7 Z_7 + c_2 Z_2 + c_3 Z_3 \,.
\ee
We claim that the soft limit corresponds to the limit $c_3,c_6\to0$. Indeed, we can look at Mandelstam variables
\be
\begin{split}
s_{12} &= x_{13}^2 = \frac{\ket{7123}}{\ket{71}\ket{23}} = \frac{c_6\ket{7623}}{\ket{23}(c_6\ket{76} + c_2\ket{72} + c_3\ket{73})}\,, \\
s_{71} &= x_{72}^2 = \frac{\ket{6712}}{\ket{67}\ket{12}} = \frac{c_3\ket{6732}}{\ket{67}(c_6\ket{62} + c_7\ket{72} + c_3\ket{32})}\,,
\end{split}
\ee
and we see that both $s_{12}$ and $s_{71}$ vanish when one takes both $c_3$ and $c_6$ to 0, as it should be in the limit $p_1\to 0$. Similarly, one can check that the other Mandelstam variables also behave as expected. Furthermore, we can look at the ratio $s_{12}/s_{71}$ in this limit,
\be
\frac{s_{12}}{s_{71}} \sim \frac{c_7 c_6\ket{67}}{c_2 c_3\ket{23}} \,.
\ee
We see that if $c_7 \ll c_2$, and if $c_3,c_6$ are taken to zero at the same rate, then $s_{12}/s_{71}\ll 1$; that is, $p_1$ is taken to 0 in a direction collinear with $p_2$. We will call this the $1||2$ \emph{soft-collinear} limit, and in this limit we have $Z_1\sim Z_2$. Similarly, when $c_2\ll c_7$ we are in the $7||1$ soft-collinear limit and we have $Z_1\sim Z_7$.

Additionally, we can adopt a six-point parametrization for the six momentum twistors $Z_2,\dots,Z_7$. Then the seven-point alphabet collapses to the six-point alphabet,
\be
u,v,w,1-u,1-v,1-w,y_u,y_v,y_w,
\ee
plus four infinitesimal letters in the seven-point cross ratios,
\be
u_5,u_6,1-u_4,1-u_7,
\ee
plus eight additional letters,
\be
x_1,x_2,x_3,x_4,x_5,x_6,x_7,x_8.
\ee
The $g$-letters are expressed as follows,
\begin{align}
g_{11} &= v,& g_{21} &= 1-v,&
g_{31} &= 1,& g_{41} &= x_5,&
g_{51} &= \frac{x_8}{y_u},&
g_{61} &= x_6,&
\nonumber\\[1ex]
g_{12} &= w,& g_{22} &= 1-w,&
g_{32} &= x_7,& g_{42} &= 1,&
g_{52} &= \frac{1}{x_8},&
g_{62} &= y_u y_v y_w,&
\nonumber\\[1ex]
g_{13} &= u,& g_{23} &= 1-u,&
g_{33} &= 1,& g_{43} &=x_3,&
g_{53} &= y_v x_8,&
g_{63} &= x_4,&
\nonumber\\[1ex]
g_{14} &= 1,& g_{24} &= 1-u_4,&
g_{34} &= 1,& g_{44} &= 1-u,&
g_{54} &= x_2,&
g_{64} &= \frac{y_u y_w}{x_8},&
\nonumber\\[1ex]
g_{15} &= u_5,& g_{25} &= 1,&
g_{35} &= 1-u,& g_{45} &= 1-v,&
g_{55} &= y_u y_v y_w,&
g_{65} &= \frac{x_8}{y_u},&
\nonumber\\[1ex]
g_{16} &= u_6,& g_{26} &= 1,&
g_{36} &= 1-v,& g_{46} &= 1-u,&
g_{56} &= \frac{1}{y_u y_v y_w},&
g_{66} &= \frac{1}{y_v x_8},&
\nonumber\\[1ex]
g_{17} &= 1,& g_{27} &= 1-u_7,&
g_{37} &= 1,& g_{47} &= 1-v,&
g_{57} &= x_1,&
g_{67} &= y_v y_w x_8.
\end{align}
The letters $1-u_4$, $1-u_7$, and $x_i$, $i=1,2,\ldots,8$ drop out of
the heptagon functions in this soft limit, and they become hexagon functions
multiplied by logarithms of $u_5$ and $u_6$.

The scaling of \eqn{eq:Z1combo} lifts to momentum supertwistors,
\be
\cZ_1 = c_6 \cZ_6 + c_7 \cZ_7 + c_2 \cZ_2 + c_3 \cZ_3 \,,
\ee
where we send $c_3,c_6\to0$ in the soft limit.

%\bibliographystyle{JHEP}
%\bibliography{heptagon}

\providecommand{\href}[2]{#2}\begingroup\raggedright\endgroup

\end{document}